\documentclass[aps, prb, preprint, amsmath,amssymb,showpacs,superscriptaddress, longbibliography]{revtex4-1}
\usepackage{graphicx}
\usepackage{ulem}
\usepackage{caption}
\usepackage{color}
\definecolor{green}{RGB}{34,139,34}
\makeatletter

\begin{document}

\title{
Rare-earth monopnoctides - family of antiferromagnets hosting magnetic Fermi arcs
} 

\author{Yevhen Kushnirenko}
\affiliation{Division of Materials Science and Engineering, Ames Laboratory, Ames, Iowa 50011, USA}
\affiliation{Department of Physics and Astronomy, Iowa State University, Ames, Iowa 50011, USA}

\author{Benjamin Schrunk}
\affiliation{Division of Materials Science and Engineering, Ames Laboratory, Ames, Iowa 50011, USA}

\author{Brinda Kuthanazhi}
\affiliation{Division of Materials Science and Engineering, Ames Laboratory, Ames, Iowa 50011, USA}
\affiliation{Department of Physics and Astronomy, Iowa State University, Ames, Iowa 50011, USA}

\author{Lin-Lin Wang}
\affiliation{Division of Materials Science and Engineering, Ames Laboratory, Ames, Iowa 50011, USA}

\author{Junyeong Ahn}
\affiliation{Department  of  Physics,  Harvard  University,  Cambridge  MA  02138,  USA}

\author{Evan O'Leary}
\affiliation{Division of Materials Science and Engineering, Ames Laboratory, Ames, Iowa 50011, USA}
\affiliation{Department of Physics and Astronomy, Iowa State University, Ames, Iowa 50011, USA}

\author{Andrew Eaton}
\affiliation{Division of Materials Science and Engineering, Ames Laboratory, Ames, Iowa 50011, USA}
\affiliation{Department of Physics and Astronomy, Iowa State University, Ames, Iowa 50011, USA}

\author{S.~L.~Bud'ko}
\affiliation{Division of Materials Science and Engineering, Ames Laboratory, Ames, Iowa 50011, USA}
\affiliation{Department of Physics and Astronomy, Iowa State University, Ames, Iowa 50011, USA}

\author{Robert-Jan Slager}
\email[]{rjs269@cam.ac.uk}
\affiliation{Department  of  Physics,  Harvard  University,  Cambridge  MA  02138,  USA}
\affiliation{TCM Group, Cavendish Laboratory, University of Cambridge, Cambridge CB3 0HE, United Kingdom}

\author{P. C. Canfield}
\email[]{canfield@ameslab.gov}
\affiliation{Division of Materials Science and Engineering, Ames Laboratory, Ames, Iowa 50011, USA}
\affiliation{Department of Physics and Astronomy, Iowa State University, Ames, Iowa 50011, USA}

\author{Adam Kaminski}
\email[]{Corresponding author: kaminski@ameslab.gov}
\affiliation{Division of Materials Science and Engineering, Ames Laboratory, Ames, Iowa 50011, USA}
\affiliation{Department of Physics and Astronomy, Iowa State University, Ames, Iowa 50011, USA}

\date{\today}

\maketitle

{\bf  Since the discovery of topological insulators a lot of research effort has been devoted to magnetic topological materials, in which non-trivial spin properties can be controlled by magnetic fields, culminating in a wealth of fundamental phenomena and possible applications. The main focus was on ferromagnetic materials that can host Weyl fermions and therefore spin textured Fermi arcs. The recent discovery of Fermi arcs and new magnetic bands splitting in antiferromagnet (AFM) NdBi has opened up new avenues for exploration. Here we show that these uncharted effects are not restricted to this specific compound, but rather emerge in CeBi, NdBi, and NdSb when they undergo paramagnetic to AFM transition. Our data show that the Fermi arcs in NdSb have 2-fold symmetry, leading to strong anisotropy that may enhance effects of spin textures on transport properties. Our findings thus demonstrate that the RBi and RSb series are materials that host magnetic Fermi arcs and may be a potential platform for modern spintronics.
}

Many of the rare-earth monopnictides RBi and RSb (where R is a rare-earth element) \cite{bartholin1979hydrostatic,Paul_Phil1992,KumigashiraPRB1996,Paul_Jcryst2001,Yun_PRB2017,Kuroda_2018} exhibit an antiferromagnetic (AFM) transition at low temperatures, while some, such as CeSb and CeBi have more than one magnetically ordered state\cite{rossat1980specific,wiener2000magnetic,Kuroda_2020,nereson_1971, nereson1972neutron, Schobinger_1973} .
Several members of this family exhibit complex evolution of electronic structure upon cascade of AFM transitions as reported by recent angle resolved photoemission spectroscopy) ARPES experiments \cite{Kuroda_2020, SakhyaNdSb2022}.
Recently, the existence of various types of topologically non-trivial states~\cite{watanabe, bouhon2021, magtqc}, including Weyl semimetal, were predicted to occur in several of these compounds \cite{DuanCommPhys2018, GuoNPJ2017, ZhuPRB2020, fang2020magnetic}. Weyl semimetals and, as a consequence, Fermi arcs that connect a pair of Weyl points were predicted to occur in ferromagnetically ordered CeBi, CeSb, and GdBi \cite{GuoNPJ2017, ZhuPRB2020, li2017predicted}. There are some experiments that seem to support this hypothesis: signatures of Weyl fermions were reported by transport measurements\cite{GuoNPJ2017} ands canning tunneling spectroscopy (STS) measurements\cite{matt2020}. However, the ferromagnetic phase of these materials can be only induced by application an external magnetic field, while in the absence of the field, these materials order only antiferromagnetically.

Since Weyl semimetals can be realized only in systems with either broken time-reversal or inversion symmetry, Fermi arcs are not expected to occur in simple AFM ordered phases such as those that were reported to occur in monopnictides by neutron diffraction experiments \cite{nereson_1971, bartholin1979hydrostatic, manfrinetti2009magnetic}.
In simple AFM phases, both symmetries are preserved, where a nonsymmorphic time-reversal symmetry has the same role as the time-reversal symmetry. In order for a AFM to realize a Weyl semimetal, it has to have a noncollinear AFM order that does not have nonsymmorphic time-reversal symmetry as in pyrochlore irridates \cite{wan2011topological}. On the other hand, a recent ARPES study on NdBi \cite{SchrunkNature2022} has shown that Fermi surface arcs emerge in the AFM phase in this material and undergo  unconventional magnetic band splitting upon cooling below T$_N$. Also, recent density functional theory (DFT) calculations \cite{wang2022multi} have shown that such Fermi surface arcs can be present in NdBi in cases of noncollinear AFM orders with multiple wave vectors (multi-q) as in 2q and 3q. In order to study how the appearance and splitting of these arcs depends on rare earth as well as pnictide we have extended our ARPES studies to RBi for R = Ce, Nd, Sm as well as to NdSb. We find that Fermi arcs exist in several members of this NaCl-structural family.

\section*{Results}

Fig. 1a shows the temperature dependence of electrical resistivity of CeBi, SmBi, NdBi, and NdSb. All curves demonstrate similar behavior: there is a high-temperature region where resistivity changes gradually, followed by a peak and a rapid decrease upon cooling. Such behavior is associated with loss of spin disorder during the AFM transition. Thermodynamic and transport data identify the following, zero applied magnetic field transition temperatures: $T_N$ = 25~K for CeBi, $T_N$ = 24~K for NdBi, $T_N$ = 15~K for NdSb, and  $T_N$ = 9~K for SmBi \cite{kuthanazhi2019metamagnetism, sakhya2021evidence, bartholin1979hydrostatic}. These are in agreement with results from other experimental techniques: scanning tunneling microscopy with a magnetic tip \cite{matt2020} and neutron scattering \cite{bartholin1979hydrostatic, manfrinetti2009magnetic, nereson_1971, nereson1972neutron, Schobinger_1973}. Whereas the NdBi and NdSb have only a single magnetic phase transition from the paramagentic phase to the AFM state, the CeBi resistivity data shows a second feature at 12.5 K. It is associated with a first order transition to a different, low temperature AFM phase. Similar situation occurs in SmBi, which also displays lower temperature AFM phase below 7K\cite{sakhya2021evidence}.  In this study, for CeBi, NdBi and NdSb, we will focus on the first AFM phase below $T_N$. For SmBi, where second transition is only 2~K below $T_N$, in which case we collected the data at 5~K to ensure we are sufficiently below $T_N$.


{\bf Fermi surface maps.} We compare the Fermi surface (FS) maps measured by ARPES in the paramagnetic and first AFM states below $T_N$ (except for SmBi, where measurements were done in second AFM state below $T_N$) in Fig. 1b-g. The Fermi surfaces of NdSb and CeBi above $T_N$, shown in Fig. 1b and c, are very similar to each other in good agreement with non-magnetic DFT calculations \cite{DuanCommPhys2018, matt2020,ZhuPRB2020}. Observed broad features are result of the 3D projection of dispersing bulk bands onto the k$_x$, k$_y$ plane of the Brillouin zone (BZ).
In the AFM state however, new sharp features appear on the Fermi surface maps (Fig. 1e and f), noticeably similar to the  previously observed NdBi elliptical, electron like pockets located near the tips of the bulk FS (Fig.1d). Another very interesting feature that emerges is a set of four disconnected contours located near the new electron-like pockets. 


The Fermi surface of CeBi in AFM state (see Fig. 1f) also demonstrates disconnected arcs very similar to that observed in NdBi and NdSb. However, there are no signs of the elliptical pockets on the map. The Fermi arcs present in NdSb and CeBi  are 2-fold symmetric (C2), while in NdBi they have 4-fold (C4) symmetry. This would be consistent with 2q magnetic ordering, as signal from (001) surface would be 4-fold symmetric and signal from (100) or (010) surface would be 2-fold symmetric. It is also possible that ARPES signal for NdSb and CeBi was originating from single (010)/(100) domain, while in NdBi case was averaged over several domains that were smaller than size our photon beam (15 $\mu$m). (See further discussion below in DFT section).

The data from SmBi measured even at the lowest magnetically ordered phase does not show any additional features in FS (Fig. 1g) in contrast to CeBi, NdBi, and NdSb. Its Fermi surface demonstrates only broad bulk features and similar to the PM state of the other materials. More data are shown in the Supplementary Information.

{\bf Band dispersion of surface states.} For the further analysis of the new states, we measured detailed data sets for the parts of the BZ where they are located. Fig. 2a shows constant-energy maps for NdSb. The elliptical pocket decreases in size and then disappears at higher binding energies indicating electron-like behavior. This behavior can also be seen from the cuts in Fig. 2b. The feature identified earlier as an arc moves out to higher k$_x$ values but remains an isolated arc and does not form closed pockets. At the same time the arc feature moves out to higher kx, essentially covering some of the region the elliptical feature had occupied. The discontinuity of this dispersion is even better seen in the cuts in Fig. 2b. For example, in the cut \#2, there are no signs of additional sharp dispersion. 
Slightly further away from $\Gamma$-point, in cut \#3, the dispersion associated with the new surface states begins to appear. Its top is still above the E$_F$ and is therefore not visible in ARPES at these low temperatures. Even further away, the band moves lower in energy, and we can see its top. In the same cut, another band with  electron-like dispersion appears. In all cuts, the lower dispersion exists in a finite momentum range and has sharp cutoffs indicating its arc character.

We performed a similar analysis for the new surface states present in NdBi. Figs. 2c and d shows constant-energy maps and the dispersion along selected cuts. There is a striking resemblance between surface state features in both materials. The elliptical pockets in NdBi also have an electron-like character. However, the band which forms them is deeper, and therefore the pockets are slightly larger. The hole-like Fermi arcs are clearly visible in these data as the sharp band exists only in a finite momentum range. Thus, the difference between NdBi and NdSb is only quantitative and not qualitative. Please note that the cuts in the NdSb are slightly skewed due to small angular misalignment of the sample during measurements.


The data for CeBi is presented in Fig. 3a, where we show the zoomed-in part of the map from Fig. 1f and several additional equal-energy plots extracted from the same data set. In Fig. 3b, we show a different data set which represents measurement on a different domain that has orthogonal orientation of the surface states. Very similar shape of the Fermi contours for two orthogonal domains indicates that the influence of matrix elements is insignificant, and both maps are adequate representations of the Fermi surface of CeBi. These maps demonstrate that arcs evolve with binding energy in a manner similar to two other materials. However, the surface electron pocket that is observed in NdBi and NdSb is absent here, most likely due to much weaker band splitting. Analysis of dispersions along cuts in certain directions reveals more details. Cut \#4 on Fig. 3c shows that the dispersion which produces the arc (marked with a white arrow) actually consists of two dispersions that are clearly split below 50 meV (see black arrows). The upper dispersion quickly loses its intensity and it becomes difficult to track its behavior beyond that point. If we move closer to the $\Gamma-X$ cut (\#5), these two dispersions merge and become indistinguishable. For the opposite direction, the upper band quickly moves higher in energy, and in cut \#3, only the bottom of this band is present. Even further, at cut \#2, both surface bands disappear, which again proves the arc character of at least lower energy branch. However, the upper dispersion does not seem to have an arc character. Most likely, it is not present in cut \#2 because, in this part of the BZ, it is located above the Fermi level. We should note that the sharp dispersion in cut \#2 
is related to the bulk states and is present in the PM states as well (see SM).  The presence of two dispersions located close to each other can also explain why the arcs on the Fermi surface of CeBi are thicker than the arcs in NdBi and NdSb. This is better seen in Fig. 1, where all maps are plotted on the same scale. Whereas we see Fermi arcs in both NdBi and NdSb, the arcs we find in CeBi are not present in CeSb\cite{Kuroda_2020}.

{\bf Temperature evolution of surface states.} The temperature evolution of surface states in CeBi across the AFM transition is shown in Fig. 3d. The spectrum on the right was measured at T = 26.5~K, which corresponds to the paramagnetic state and reveals no traces of the surface states with only broad 3D bands being present. At 24~K the surface states are already clearly seen. This result is in agreement with $T_N$ = 25~K obtained from our resistivity measurements and other methods. Upon further cooling down to 13 K, intensity of the surface states increase. We limit the lower temperature to 13~K in order to compare temperature dependent data in the initial AFM ordered state upon cooling from the paramagnetic state.

We compare the temperature evolution of the surface states for NdSb and NdBi in Fig. 4. The Fermi surface maps measured at several different temperatures are shown in panels a and d, horizontal cuts in panels b and e and vertical cuts in panels c and f, respectively. The Fermi surface maps for both NdSb and NdBi (Fig. 4 a and d) show that the separation between electron-like pockets and hole like Fermi arcs decreases with increased temperature. At the same time, the intensity of the electron-like pocket decreases. Similar behavior is seen in band dispersion plots shown in Fig. 4 b, c, e, f. At lowest temperature the hole and electron bands are well separated. This separation decreases upon increasing temperature, with electron band changing its energy more than the hole band. At temperature close to the transition temperature, both bands merge along the horizontal cut, while the intensity of the electron band becomes very weak. Finally, at T= 15.5~K for NdSb and 25~K for NdBi, all features associated with the surface state disappear in both Fermi surface maps and dispersion. 

The comparison of data measured at similar effective temperatures T$_{}$=(T/T$_N$) reveals significant differences between CeBi and NdBi/NdSb. For example, the surface state bands in NdBi at 15K (T$_{}$=0.6) shown in Fig. 4e are already substantially split, while in CeBi, at lowest temperature shown in Fig. 3c and d, i. e. for even lower (T$_{eff}$=0.54) the splitting is much smaller and visible only in certain cuts. One would have to compare the NdBi data measured  at 20~K with CeBi data measured at 13~K to find some  similarity. In NdBi at T=20~K, the arc merges with a part of the electron pocket, forming one broad feature on the Fermi Surface similar to one observed in CeBi. At the same time, the rest of the electron pocket tends to lose all its intensity at high temperatures. Also, the behavior of the surface state dispersions on the high-temperature spectrum in Fig.4e is reminiscent of the behavior in Fig.3c \#4, while the high-temperature spectrum in Fig.4f is similar to Fig. 3e. The spectrum measured at T= 20~K in NdBi shows that the bottom of the electron dispersion is drastically more intense than the rest of it. In CeBi, we see a similar situation: above the band associated with the arc, we see an additional feature, which we associate with the bottom of the electron dispersion. All these similarities indicate that in CeBi, electron-like SS dispersion is indeed present, but it is not fully developed due to much smaller splitting. This is likely due to weaker effective magnetic moment. Furthermore, first order transition to a different AFM phase that occurs at 12.5~K, prevents full formation of this feature that is expected to occur at lower temperatures, as it occurs in NdBi and NdSb.

{\bf DFT calculations.} Previously, we have proposed \cite{wang2022multi} that noncollinear multi-$q$ AFM structures with two and three wave vectors (2$q$ and 3$q$) can give arise to unconventional surface state pairs that match well with the surface hole Fermi arcs and electron pockets as observed in NdBi below $T_N$\cite{SchrunkNature2022}. Here we extend such multi-$q$ DFT calculation and analysis to NdSb. Figure 5a and b show the orientation of magnetic moments in 2$q$ NdSb and the corresponding 2D Fermi surface on (010) side surface in the full surface BZ. The existence of the unconventional surface state pairs only along the $\Gamma-X$ not $\Gamma-Y$ direction agrees well with the ARPES data in Fig.1d. With a shift of $E_F$ by 30 meV, the curvature of the hole arc connecting to bulk band and the elliptical surface state electron pocket match well with the ARPES data, except for the cusp of the latter enveloping more with bulk bands, which is due to a larger projection of the bulk bands in the calculation. As zoomed in Fig.5c at different binding energies, the unconventional surface state pairs also compare well to those in Fig.2a, indicating their hole and electron character, respectively. Importantly, the in-plane spin textures are plotted in Fig.5d for the corresponding surface states in Fig.5c. As shown by the orange arrows, the surface hole arc and electron pocket have opposite electron spins, which are similar to those in NdBi \cite{wang2022multi}, despite Sb is lighter than Bi. Furthermore, DFT calculation indicates that these unconventional surface pairs reside in the band-folding bulk gap as shown in Fig.5e by cut \#1 along the $\Gamma-X$ direction. The other cuts show the dispersion of the unconventional surface pairs in other area also match well with the ARPES data in Fig.2b.  

We did not consider 3q ordering for NdSb, since, in these cases, the calculated FS has C4 symmetry that does not agree with our ARPES results. It is possible and likely that NdBi also has 2q ordering. In this case, we can observe 4-fold symmetric FS in the experiment at the (001) surface (see Fig. 5a) or if the domain sizes on (010)/(100) surfaces in NdBi are smaller than our laser beam ($\sim$15 $\mu$m). The 4-fold symmetric FS could be observed in the latter case because the signal would be a superposition of signals from domains of two different, perpendicular orientations.

In the future, more theoretical modeling is required to explain the intriguing temperature-dependent splitting of these spin-textured surface states.

\section*{Conclusions}
In this work, we show that the exciting novel breakthrough features discovered in NdBi, such as spontaneous Fermi surface generation and anomalous magnetic splitting effects that culminate in effective manipulable Fermi Arcs, are not restricted to that specific compound. Rather they pinpoint to a general and uncharted mechanism that follows a definitive trend over a wide range of mono-pnictide family members when they undergo AFM transitions. This trend, in terms of relative intensity of the Fermi arcs and energy separation of the magenetic band splitting seems to scale with the magnetic moments of the rare-earth elements, with Nd having largest magnetic moment and strongest effects amongst compounds we studied. Sm has smallest magnetic moment with SmBi  not exhibiting Fermi arcs nor splitting. Ce with its magnetic in between exhibits modest Fermi arcs and very small magnetic splitting. As such our results not only set a benchmark for the investigation of novel magnetic surface states in a experimentally accessible class of materials, but open possibility that other antiferromagnets can also host similar spin textured Fermi arcs and magnetic splitting. 


\section*{References}
\bibliography{NdBi_arcs}

\begin{thebibliography}{44}%
\makeatletter
\providecommand \@ifxundefined [1]{%
 \@ifx{#1\undefined}
}%
\providecommand \@ifnum [1]{%
 \ifnum #1\expandafter \@firstoftwo
 \else \expandafter \@secondoftwo
 \fi
}%
\providecommand \@ifx [1]{%
 \ifx #1\expandafter \@firstoftwo
 \else \expandafter \@secondoftwo
 \fi
}%
\providecommand \natexlab [1]{#1}%
\providecommand \enquote  [1]{``#1''}%
\providecommand \bibnamefont  [1]{#1}%
\providecommand \bibfnamefont [1]{#1}%
\providecommand \citenamefont [1]{#1}%
\providecommand \href@noop [0]{\@secondoftwo}%
\providecommand \href [0]{\begingroup \@sanitize@url \@href}%
\providecommand \@href[1]{\@@startlink{#1}\@@href}%
\providecommand \@@href[1]{\endgroup#1\@@endlink}%
\providecommand \@sanitize@url [0]{\catcode `\\12\catcode `\$12\catcode
  `\&12\catcode `\#12\catcode `\^12\catcode `\_12\catcode `\%12\relax}%
\providecommand \@@startlink[1]{}%
\providecommand \@@endlink[0]{}%
\providecommand \url  [0]{\begingroup\@sanitize@url \@url }%
\providecommand \@url [1]{\endgroup\@href {#1}{\urlprefix }}%
\providecommand \urlprefix  [0]{URL }%
\providecommand \Eprint [0]{\href }%
\providecommand \doibase [0]{http://dx.doi.org/}%
\providecommand \selectlanguage [0]{\@gobble}%
\providecommand \bibinfo  [0]{\@secondoftwo}%
\providecommand \bibfield  [0]{\@secondoftwo}%
\providecommand \translation [1]{[#1]}%
\providecommand \BibitemOpen [0]{}%
\providecommand \bibitemStop [0]{}%
\providecommand \bibitemNoStop [0]{.\EOS\space}%
\providecommand \EOS [0]{\spacefactor3000\relax}%
\providecommand \BibitemShut  [1]{\csname bibitem#1\endcsname}%
\let\auto@bib@innerbib\@empty
\bibitem [{\citenamefont {Bartholin}\ \emph {et~al.}(1979)\citenamefont
  {Bartholin}, \citenamefont {Burlet}, \citenamefont {Quezel}, \citenamefont
  {Rossat-Mignod},\ and\ \citenamefont {Vogt}}]{bartholin1979hydrostatic}%
  \BibitemOpen
  \bibfield  {author} {\bibinfo {author} {\bibfnamefont {H}~\bibnamefont
  {Bartholin}}, \bibinfo {author} {\bibfnamefont {P}~\bibnamefont {Burlet}},
  \bibinfo {author} {\bibfnamefont {S}~\bibnamefont {Quezel}}, \bibinfo
  {author} {\bibfnamefont {J}~\bibnamefont {Rossat-Mignod}}, \ and\ \bibinfo
  {author} {\bibfnamefont {O}~\bibnamefont {Vogt}},\ }\bibfield  {title}
  {\enquote {\bibinfo {title} {Hydrostatic pressure effects and neutron
  diffraction studies of cebi phase diagram},}\ }\href@noop {} {\bibfield
  {journal} {\bibinfo  {journal} {Le Journal de Physique Colloques}\ }\textbf
  {\bibinfo {volume} {40}},\ \bibinfo {pages} {C5--130} (\bibinfo {year}
  {1979})}\BibitemShut {NoStop}%
\bibitem [{\citenamefont {Canfield}\ and\ \citenamefont
  {Fisk}(1992)}]{Paul_Phil1992}%
  \BibitemOpen
  \bibfield  {author} {\bibinfo {author} {\bibfnamefont {P.~C.}\ \bibnamefont
  {Canfield}}\ and\ \bibinfo {author} {\bibfnamefont {Z}~\bibnamefont {Fisk}},\
  }\bibfield  {title} {\enquote {\bibinfo {title} {{Growth of single crystals
  from metallic fluxes}},}\ }\href@noop {} {\bibfield  {journal} {\bibinfo
  {journal} {Philosophical Magazine Part B}\ }\textbf {\bibinfo {volume}
  {65}},\ \bibinfo {pages} {1117--1123} (\bibinfo {year} {1992})}\BibitemShut
  {NoStop}%
\bibitem [{\citenamefont {Kumigashira}\ \emph {et~al.}(1996)\citenamefont
  {Kumigashira}, \citenamefont {Yang}, \citenamefont {Yokoya}, \citenamefont
  {Chainani}, \citenamefont {Takahashi}, \citenamefont {Uesawa}, \citenamefont
  {Suzuki}, \citenamefont {Sakai},\ and\ \citenamefont
  {Kaneta}}]{KumigashiraPRB1996}%
  \BibitemOpen
  \bibfield  {author} {\bibinfo {author} {\bibfnamefont {H.}~\bibnamefont
  {Kumigashira}}, \bibinfo {author} {\bibfnamefont {S.-H.}\ \bibnamefont
  {Yang}}, \bibinfo {author} {\bibfnamefont {T.}~\bibnamefont {Yokoya}},
  \bibinfo {author} {\bibfnamefont {A.}~\bibnamefont {Chainani}}, \bibinfo
  {author} {\bibfnamefont {T.}~\bibnamefont {Takahashi}}, \bibinfo {author}
  {\bibfnamefont {A.}~\bibnamefont {Uesawa}}, \bibinfo {author} {\bibfnamefont
  {T.}~\bibnamefont {Suzuki}}, \bibinfo {author} {\bibfnamefont
  {O.}~\bibnamefont {Sakai}}, \ and\ \bibinfo {author} {\bibfnamefont
  {Y.}~\bibnamefont {Kaneta}},\ }\bibfield  {title} {\enquote {\bibinfo {title}
  {High-resolution angle-resolved photoemission spectroscopy of cebi},}\ }\href
  {\doibase 10.1103/PhysRevB.54.9341} {\bibfield  {journal} {\bibinfo
  {journal} {Phys. Rev. B}\ }\textbf {\bibinfo {volume} {54}},\ \bibinfo
  {pages} {9341--9345} (\bibinfo {year} {1996})}\BibitemShut {NoStop}%
\bibitem [{\citenamefont {Canfield}\ and\ \citenamefont
  {Fisher}(2001)}]{Paul_Jcryst2001}%
  \BibitemOpen
  \bibfield  {author} {\bibinfo {author} {\bibfnamefont {P.~C.}\ \bibnamefont
  {Canfield}}\ and\ \bibinfo {author} {\bibfnamefont {Ian~R}\ \bibnamefont
  {Fisher}},\ }\bibfield  {title} {\enquote {\bibinfo {title}
  {{High-temperature solution growth of intermetallic single crystals and
  quasicrystals}},}\ }\href@noop {} {\bibfield  {journal} {\bibinfo  {journal}
  {Journal of Crystal Growth}\ }\textbf {\bibinfo {volume} {225}},\ \bibinfo
  {pages} {155--161} (\bibinfo {year} {2001})}\BibitemShut {NoStop}%
\bibitem [{\citenamefont {Wu}\ \emph {et~al.}(2017)\citenamefont {Wu},
  \citenamefont {Lee}, \citenamefont {Kong}, \citenamefont {Mou}, \citenamefont
  {Jiang}, \citenamefont {Huang}, \citenamefont {Bud'ko}, \citenamefont
  {Canfield},\ and\ \citenamefont {Kaminski}}]{Yun_PRB2017}%
  \BibitemOpen
  \bibfield  {author} {\bibinfo {author} {\bibfnamefont {Yun}\ \bibnamefont
  {Wu}}, \bibinfo {author} {\bibfnamefont {Yongbin}\ \bibnamefont {Lee}},
  \bibinfo {author} {\bibfnamefont {Tai}\ \bibnamefont {Kong}}, \bibinfo
  {author} {\bibfnamefont {Dai-xiang}\ \bibnamefont {Mou}}, \bibinfo {author}
  {\bibfnamefont {Rui}\ \bibnamefont {Jiang}}, \bibinfo {author} {\bibfnamefont
  {Lunan}\ \bibnamefont {Huang}}, \bibinfo {author} {\bibfnamefont {Sergey~L}\
  \bibnamefont {Bud'ko}}, \bibinfo {author} {\bibfnamefont {P.~C.}\
  \bibnamefont {Canfield}}, \ and\ \bibinfo {author} {\bibfnamefont
  {A}~\bibnamefont {Kaminski}},\ }\bibfield  {title} {\enquote {\bibinfo
  {title} {{Electronic structure of RSb (R=Y, Ce, Gd, Dy, Ho, Tm, Lu) studied
  by angle-resolved photoemission spectroscopy}},}\ }\href@noop {} {\bibfield
  {journal} {\bibinfo  {journal} {Phys. Rev. B}\ }\textbf {\bibinfo {volume}
  {96}},\ \bibinfo {pages} {035134} (\bibinfo {year} {2017})}\BibitemShut
  {NoStop}%
\bibitem [{\citenamefont {Kuroda}\ \emph {et~al.}(2018)\citenamefont {Kuroda},
  \citenamefont {Ochi}, \citenamefont {Suzuki}, \citenamefont {Hirayama},
  \citenamefont {Nakayama}, \citenamefont {Noguchi}, \citenamefont {Bareille},
  \citenamefont {Akebi}, \citenamefont {Kunisada}, \citenamefont {Muro},
  \citenamefont {Watson}, \citenamefont {Kitazawa}, \citenamefont {Haga},
  \citenamefont {Kim}, \citenamefont {Hoesch}, \citenamefont {Shin},
  \citenamefont {Arita},\ and\ \citenamefont {Kondo}}]{Kuroda_2018}%
  \BibitemOpen
  \bibfield  {author} {\bibinfo {author} {\bibfnamefont {Kenta}\ \bibnamefont
  {Kuroda}}, \bibinfo {author} {\bibfnamefont {M.}~\bibnamefont {Ochi}},
  \bibinfo {author} {\bibfnamefont {H.~S.}\ \bibnamefont {Suzuki}}, \bibinfo
  {author} {\bibfnamefont {M.}~\bibnamefont {Hirayama}}, \bibinfo {author}
  {\bibfnamefont {M.}~\bibnamefont {Nakayama}}, \bibinfo {author}
  {\bibfnamefont {R.}~\bibnamefont {Noguchi}}, \bibinfo {author} {\bibfnamefont
  {C.}~\bibnamefont {Bareille}}, \bibinfo {author} {\bibfnamefont
  {S.}~\bibnamefont {Akebi}}, \bibinfo {author} {\bibfnamefont
  {S.}~\bibnamefont {Kunisada}}, \bibinfo {author} {\bibfnamefont
  {T.}~\bibnamefont {Muro}}, \bibinfo {author} {\bibfnamefont {M.~D.}\
  \bibnamefont {Watson}}, \bibinfo {author} {\bibfnamefont {H.}~\bibnamefont
  {Kitazawa}}, \bibinfo {author} {\bibfnamefont {Y.}~\bibnamefont {Haga}},
  \bibinfo {author} {\bibfnamefont {T.~K.}\ \bibnamefont {Kim}}, \bibinfo
  {author} {\bibfnamefont {M.}~\bibnamefont {Hoesch}}, \bibinfo {author}
  {\bibfnamefont {S.}~\bibnamefont {Shin}}, \bibinfo {author} {\bibfnamefont
  {R.}~\bibnamefont {Arita}}, \ and\ \bibinfo {author} {\bibfnamefont
  {Takeshi}\ \bibnamefont {Kondo}},\ }\bibfield  {title} {\enquote {\bibinfo
  {title} {Experimental determination of the topological phase diagram in
  cerium monopnictides},}\ }\href {\doibase 10.1103/PhysRevLett.120.086402}
  {\bibfield  {journal} {\bibinfo  {journal} {Phys. Rev. Lett.}\ }\textbf
  {\bibinfo {volume} {120}},\ \bibinfo {pages} {086402} (\bibinfo {year}
  {2018})}\BibitemShut {NoStop}%
\bibitem [{\citenamefont {Rossat-Mignod}\ \emph {et~al.}(1980)\citenamefont
  {Rossat-Mignod}, \citenamefont {Burlet}, \citenamefont {Bartholin},
  \citenamefont {Vogt},\ and\ \citenamefont {Lagnier}}]{rossat1980specific}%
  \BibitemOpen
  \bibfield  {author} {\bibinfo {author} {\bibfnamefont {J}~\bibnamefont
  {Rossat-Mignod}}, \bibinfo {author} {\bibfnamefont {P}~\bibnamefont
  {Burlet}}, \bibinfo {author} {\bibfnamefont {H}~\bibnamefont {Bartholin}},
  \bibinfo {author} {\bibfnamefont {O}~\bibnamefont {Vogt}}, \ and\ \bibinfo
  {author} {\bibfnamefont {R}~\bibnamefont {Lagnier}},\ }\bibfield  {title}
  {\enquote {\bibinfo {title} {Specific heat analysis of the magnetic phase
  diagram of cesb},}\ }\href@noop {} {\bibfield  {journal} {\bibinfo  {journal}
  {Journal of Physics C: Solid State Physics}\ }\textbf {\bibinfo {volume}
  {13}},\ \bibinfo {pages} {6381} (\bibinfo {year} {1980})}\BibitemShut
  {NoStop}%
\bibitem [{\citenamefont {Wiener}\ and\ \citenamefont
  {Canfield}(2000)}]{wiener2000magnetic}%
  \BibitemOpen
  \bibfield  {author} {\bibinfo {author} {\bibfnamefont {TA}~\bibnamefont
  {Wiener}}\ and\ \bibinfo {author} {\bibfnamefont {PC}~\bibnamefont
  {Canfield}},\ }\bibfield  {title} {\enquote {\bibinfo {title} {Magnetic phase
  diagram of flux-grown single crystals of cesb},}\ }\href@noop {} {\bibfield
  {journal} {\bibinfo  {journal} {Journal of Alloys and Compounds}\ }\textbf
  {\bibinfo {volume} {303}},\ \bibinfo {pages} {505--508} (\bibinfo {year}
  {2000})}\BibitemShut {NoStop}%
\bibitem [{\citenamefont {Kuroda}\ \emph {et~al.}(2020)\citenamefont {Kuroda},
  \citenamefont {Arai}, \citenamefont {Rezaei}, \citenamefont {Kunisada},
  \citenamefont {Sakuragi}, \citenamefont {Alaei}, \citenamefont {Kinoshita},
  \citenamefont {Bareille}, \citenamefont {Noguchi}, \citenamefont {Nakayama},
  \citenamefont {Akebi}, \citenamefont {Sakano}, \citenamefont {Kawaguchi},
  \citenamefont {Arita}, \citenamefont {Ideta}, \citenamefont {Tanaka},
  \citenamefont {Kitazawa}, \citenamefont {Okazaki}, \citenamefont {Tokunaga},
  \citenamefont {Haga}, \citenamefont {Shin}, \citenamefont {Suzuki},
  \citenamefont {Arita},\ and\ \citenamefont {Kondo}}]{Kuroda_2020}%
  \BibitemOpen
  \bibfield  {author} {\bibinfo {author} {\bibfnamefont {Kenta}\ \bibnamefont
  {Kuroda}}, \bibinfo {author} {\bibfnamefont {Y}~\bibnamefont {Arai}},
  \bibinfo {author} {\bibfnamefont {N}~\bibnamefont {Rezaei}}, \bibinfo
  {author} {\bibfnamefont {S}~\bibnamefont {Kunisada}}, \bibinfo {author}
  {\bibfnamefont {S}~\bibnamefont {Sakuragi}}, \bibinfo {author} {\bibfnamefont
  {M}~\bibnamefont {Alaei}}, \bibinfo {author} {\bibfnamefont {Y}~\bibnamefont
  {Kinoshita}}, \bibinfo {author} {\bibfnamefont {C}~\bibnamefont {Bareille}},
  \bibinfo {author} {\bibfnamefont {R}~\bibnamefont {Noguchi}}, \bibinfo
  {author} {\bibfnamefont {M}~\bibnamefont {Nakayama}}, \bibinfo {author}
  {\bibfnamefont {S}~\bibnamefont {Akebi}}, \bibinfo {author} {\bibfnamefont
  {M}~\bibnamefont {Sakano}}, \bibinfo {author} {\bibfnamefont {K}~\bibnamefont
  {Kawaguchi}}, \bibinfo {author} {\bibfnamefont {M}~\bibnamefont {Arita}},
  \bibinfo {author} {\bibfnamefont {S}~\bibnamefont {Ideta}}, \bibinfo {author}
  {\bibfnamefont {K}~\bibnamefont {Tanaka}}, \bibinfo {author} {\bibfnamefont
  {H}~\bibnamefont {Kitazawa}}, \bibinfo {author} {\bibfnamefont
  {K}~\bibnamefont {Okazaki}}, \bibinfo {author} {\bibfnamefont
  {M}~\bibnamefont {Tokunaga}}, \bibinfo {author} {\bibfnamefont
  {Y}~\bibnamefont {Haga}}, \bibinfo {author} {\bibfnamefont {Shik}\
  \bibnamefont {Shin}}, \bibinfo {author} {\bibfnamefont {H~S}\ \bibnamefont
  {Suzuki}}, \bibinfo {author} {\bibfnamefont {R}~\bibnamefont {Arita}}, \ and\
  \bibinfo {author} {\bibfnamefont {Takeshi}\ \bibnamefont {Kondo}},\
  }\bibfield  {title} {\enquote {\bibinfo {title} {{Devil's staircase
  transition of the electronic structures in CeSb}},}\ }\href@noop {}
  {\bibfield  {journal} {\bibinfo  {journal} {Nature Communications}\ }\textbf
  {\bibinfo {volume} {11}},\ \bibinfo {pages} {1--9} (\bibinfo {year}
  {2020})}\BibitemShut {NoStop}%
\bibitem [{\citenamefont {Nereson}\ and\ \citenamefont
  {Arnold}(1971)}]{nereson_1971}%
  \BibitemOpen
  \bibfield  {author} {\bibinfo {author} {\bibfnamefont {N.}~\bibnamefont
  {Nereson}}\ and\ \bibinfo {author} {\bibfnamefont {G.}~\bibnamefont
  {Arnold}},\ }\bibfield  {title} {\enquote {\bibinfo {title} {Magnetic
  properties of cebi, ndbi, tbbi, and dybi},}\ }\href {\doibase
  10.1063/1.1660369} {\bibfield  {journal} {\bibinfo  {journal} {Journal of
  Applied Physics}\ }\textbf {\bibinfo {volume} {42}},\ \bibinfo {pages}
  {1625--1627} (\bibinfo {year} {1971})},\ \Eprint
  {http://arxiv.org/abs/https://doi.org/10.1063/1.1660369}
  {https://doi.org/10.1063/1.1660369} \BibitemShut {NoStop}%
\bibitem [{\citenamefont {Nereson}\ and\ \citenamefont
  {Struebing}(1972)}]{nereson1972neutron}%
  \BibitemOpen
  \bibfield  {author} {\bibinfo {author} {\bibfnamefont {Norris}\ \bibnamefont
  {Nereson}}\ and\ \bibinfo {author} {\bibfnamefont {Vernon}\ \bibnamefont
  {Struebing}},\ }\bibfield  {title} {\enquote {\bibinfo {title} {Neutron
  diffraction studies on dysb, ndsb, and cesb},}\ }in\ \href@noop {} {\emph
  {\bibinfo {booktitle} {AIP Conference Proceedings}}},\ Vol.~\bibinfo {volume}
  {5}\ (\bibinfo {organization} {American Institute of Physics},\ \bibinfo
  {year} {1972})\ pp.\ \bibinfo {pages} {1385--1389}\BibitemShut {NoStop}%
\bibitem [{\citenamefont {Schobinger-Papamantellos}\ \emph
  {et~al.}(1973)\citenamefont {Schobinger-Papamantellos}, \citenamefont
  {Fischer}, \citenamefont {Vogt},\ and\ \citenamefont
  {Kaldis}}]{Schobinger_1973}%
  \BibitemOpen
  \bibfield  {author} {\bibinfo {author} {\bibfnamefont {P}~\bibnamefont
  {Schobinger-Papamantellos}}, \bibinfo {author} {\bibfnamefont
  {P}~\bibnamefont {Fischer}}, \bibinfo {author} {\bibfnamefont
  {O}~\bibnamefont {Vogt}}, \ and\ \bibinfo {author} {\bibfnamefont
  {E}~\bibnamefont {Kaldis}},\ }\bibfield  {title} {\enquote {\bibinfo {title}
  {Magnetic ordering of neodymium monopnictides determined by neutron
  diffraction},}\ }\href {\doibase 10.1088/0022-3719/6/4/020} {\bibfield
  {journal} {\bibinfo  {journal} {Journal of Physics C: Solid State Physics}\
  }\textbf {\bibinfo {volume} {6}},\ \bibinfo {pages} {725--737} (\bibinfo
  {year} {1973})}\BibitemShut {NoStop}%
\bibitem [{\citenamefont {Sakhya}\ \emph {et~al.}(2022)\citenamefont {Sakhya},
  \citenamefont {Wang}, \citenamefont {Kabir}, \citenamefont {Huang},
  \citenamefont {Hosen}, \citenamefont {Singh}, \citenamefont {Regmi},
  \citenamefont {Dhakal}, \citenamefont {Dimitri}, \citenamefont {Sprague},
  \citenamefont {Smith}, \citenamefont {Bauer}, \citenamefont {Ronning},
  \citenamefont {Bansil},\ and\ \citenamefont {Neupane}}]{SakhyaNdSb2022}%
  \BibitemOpen
  \bibfield  {author} {\bibinfo {author} {\bibfnamefont {Anup~Pradhan}\
  \bibnamefont {Sakhya}}, \bibinfo {author} {\bibfnamefont {Baokai}\
  \bibnamefont {Wang}}, \bibinfo {author} {\bibfnamefont {Firoza}\ \bibnamefont
  {Kabir}}, \bibinfo {author} {\bibfnamefont {Cheng-Yi}\ \bibnamefont {Huang}},
  \bibinfo {author} {\bibfnamefont {M.~Mofazzel}\ \bibnamefont {Hosen}},
  \bibinfo {author} {\bibfnamefont {Bahadur}\ \bibnamefont {Singh}}, \bibinfo
  {author} {\bibfnamefont {Sabin}\ \bibnamefont {Regmi}}, \bibinfo {author}
  {\bibfnamefont {Gyanendra}\ \bibnamefont {Dhakal}}, \bibinfo {author}
  {\bibfnamefont {Klauss}\ \bibnamefont {Dimitri}}, \bibinfo {author}
  {\bibfnamefont {Milo}\ \bibnamefont {Sprague}}, \bibinfo {author}
  {\bibfnamefont {Robert}\ \bibnamefont {Smith}}, \bibinfo {author}
  {\bibfnamefont {Eric~D.}\ \bibnamefont {Bauer}}, \bibinfo {author}
  {\bibfnamefont {Filip}\ \bibnamefont {Ronning}}, \bibinfo {author}
  {\bibfnamefont {Arun}\ \bibnamefont {Bansil}}, \ and\ \bibinfo {author}
  {\bibfnamefont {Madhab}\ \bibnamefont {Neupane}},\ }\bibfield  {title}
  {\enquote {\bibinfo {title} {Complex electronic structure evolution of ndsb
  across the magnetic transition},}\ }\href {https://arxiv.org/abs/2203.05879}
  {\bibfield  {journal} {\bibinfo  {journal} {arxiv:2203.05879}\ } (\bibinfo
  {year} {2022})}\BibitemShut {NoStop}%
\bibitem [{\citenamefont {Watanabe}\ \emph {et~al.}(2018)\citenamefont
  {Watanabe}, \citenamefont {Po},\ and\ \citenamefont {Vishwanath}}]{watanabe}%
  \BibitemOpen
  \bibfield  {author} {\bibinfo {author} {\bibfnamefont {Haruki}\ \bibnamefont
  {Watanabe}}, \bibinfo {author} {\bibfnamefont {Hoi~Chun}\ \bibnamefont {Po}},
  \ and\ \bibinfo {author} {\bibfnamefont {Ashvin}\ \bibnamefont
  {Vishwanath}},\ }\bibfield  {title} {\enquote {\bibinfo {title} {Structure
  and topology of band structures in the 1651 magnetic space groups},}\ }\href
  {\doibase 10.1126/sciadv.aat8685} {\bibfield  {journal} {\bibinfo  {journal}
  {Science Advances}\ }\textbf {\bibinfo {volume} {4}},\ \bibinfo {pages}
  {eaat8685} (\bibinfo {year} {2018})}\BibitemShut {NoStop}%
\bibitem [{\citenamefont {Bouhon}\ \emph {et~al.}(2021)\citenamefont {Bouhon},
  \citenamefont {Lange},\ and\ \citenamefont {Slager}}]{bouhon2021}%
  \BibitemOpen
  \bibfield  {author} {\bibinfo {author} {\bibfnamefont {Adrien}\ \bibnamefont
  {Bouhon}}, \bibinfo {author} {\bibfnamefont {Gunnar~F.}\ \bibnamefont
  {Lange}}, \ and\ \bibinfo {author} {\bibfnamefont {Robert-Jan}\ \bibnamefont
  {Slager}},\ }\bibfield  {title} {\enquote {\bibinfo {title} {Topological
  correspondence between magnetic space group representations and
  subdimensions},}\ }\href {\doibase 10.1103/PhysRevB.103.245127} {\bibfield
  {journal} {\bibinfo  {journal} {Phys. Rev. B}\ }\textbf {\bibinfo {volume}
  {103}},\ \bibinfo {pages} {245127} (\bibinfo {year} {2021})}\BibitemShut
  {NoStop}%
\bibitem [{\citenamefont {Elcoro}\ \emph {et~al.}(2021)\citenamefont {Elcoro},
  \citenamefont {Wieder}, \citenamefont {Song}, \citenamefont {Xu},
  \citenamefont {Bradlyn},\ and\ \citenamefont {Bernevig}}]{magtqc}%
  \BibitemOpen
  \bibfield  {author} {\bibinfo {author} {\bibfnamefont {Luis}\ \bibnamefont
  {Elcoro}}, \bibinfo {author} {\bibfnamefont {Benjamin~J.}\ \bibnamefont
  {Wieder}}, \bibinfo {author} {\bibfnamefont {Zhida}\ \bibnamefont {Song}},
  \bibinfo {author} {\bibfnamefont {Yuanfeng}\ \bibnamefont {Xu}}, \bibinfo
  {author} {\bibfnamefont {Barry}\ \bibnamefont {Bradlyn}}, \ and\ \bibinfo
  {author} {\bibfnamefont {B.~Andrei}\ \bibnamefont {Bernevig}},\ }\bibfield
  {title} {\enquote {\bibinfo {title} {Magnetic topological quantum
  chemistry},}\ }\href {\doibase 10.1038/s41467-021-26241-8} {\bibfield
  {journal} {\bibinfo  {journal} {Nature Communications}\ }\textbf {\bibinfo
  {volume} {12}},\ \bibinfo {pages} {5965} (\bibinfo {year}
  {2021})}\BibitemShut {NoStop}%
\bibitem [{\citenamefont {{Duan, Xu}}\ \emph {et~al.}(2018)\citenamefont
  {{Duan, Xu}}, \citenamefont {{Wu, Fan}}, \citenamefont {{Chen, Jia}},
  \citenamefont {{Zhang, Peiran}}, \citenamefont {{Liu, Yang}}, \citenamefont
  {{Yuan, Huiqiu}},\ and\ \citenamefont {{Cao, Chao}}}]{DuanCommPhys2018}%
  \BibitemOpen
  \bibfield  {author} {\bibinfo {author} {\bibnamefont {{Duan, Xu}}}, \bibinfo
  {author} {\bibnamefont {{Wu, Fan}}}, \bibinfo {author} {\bibnamefont {{Chen,
  Jia}}}, \bibinfo {author} {\bibnamefont {{Zhang, Peiran}}}, \bibinfo {author}
  {\bibnamefont {{Liu, Yang}}}, \bibinfo {author} {\bibnamefont {{Yuan,
  Huiqiu}}}, \ and\ \bibinfo {author} {\bibnamefont {{Cao, Chao}}},\ }\bibfield
   {title} {\enquote {\bibinfo {title} {{Tunable electronic structure and
  topological properties of LnPn (Ln=Ce, Pr, Sm, Gd, Yb; Pn=Sb, Bi)}},}\
  }\href@noop {} {\bibfield  {journal} {\bibinfo  {journal} {Communications
  Physics}\ }\textbf {\bibinfo {volume} {1}},\ \bibinfo {pages} {71} (\bibinfo
  {year} {2018})}\BibitemShut {NoStop}%
\bibitem [{\citenamefont {Guo}\ \emph {et~al.}(2017)\citenamefont {Guo},
  \citenamefont {Cao}, \citenamefont {Smidman}, \citenamefont {Wu},
  \citenamefont {Zhang}, \citenamefont {Steglich}, \citenamefont {Zhang},\ and\
  \citenamefont {Yuan}}]{GuoNPJ2017}%
  \BibitemOpen
  \bibfield  {author} {\bibinfo {author} {\bibfnamefont {Chunyu}\ \bibnamefont
  {Guo}}, \bibinfo {author} {\bibfnamefont {Chao}\ \bibnamefont {Cao}},
  \bibinfo {author} {\bibfnamefont {Michael}\ \bibnamefont {Smidman}}, \bibinfo
  {author} {\bibfnamefont {Fan}\ \bibnamefont {Wu}}, \bibinfo {author}
  {\bibfnamefont {Yongjun}\ \bibnamefont {Zhang}}, \bibinfo {author}
  {\bibfnamefont {Frank}\ \bibnamefont {Steglich}}, \bibinfo {author}
  {\bibfnamefont {Fu-Chun}\ \bibnamefont {Zhang}}, \ and\ \bibinfo {author}
  {\bibfnamefont {Huiqiu}\ \bibnamefont {Yuan}},\ }\bibfield  {title} {\enquote
  {\bibinfo {title} {{Possible Weyl fermions in the magnetic Kondo system
  CeSb}},}\ }\href@noop {} {\bibfield  {journal} {\bibinfo  {journal} {Npj
  Quantum Materials}\ }\textbf {\bibinfo {volume} {2}},\ \bibinfo {pages} {--6}
  (\bibinfo {year} {2017})}\BibitemShut {NoStop}%
\bibitem [{\citenamefont {Huang}\ \emph {et~al.}(2020)\citenamefont {Huang},
  \citenamefont {Lane}, \citenamefont {Cao}, \citenamefont {Zhi}, \citenamefont
  {Liu}, \citenamefont {Matt}, \citenamefont {Kuthanazhi}, \citenamefont
  {Canfield}, \citenamefont {Yarotski}, \citenamefont {Taylor},\ and\
  \citenamefont {Zhu}}]{ZhuPRB2020}%
  \BibitemOpen
  \bibfield  {author} {\bibinfo {author} {\bibfnamefont {Zhao}\ \bibnamefont
  {Huang}}, \bibinfo {author} {\bibfnamefont {Christopher}\ \bibnamefont
  {Lane}}, \bibinfo {author} {\bibfnamefont {Chao}\ \bibnamefont {Cao}},
  \bibinfo {author} {\bibfnamefont {Guo-Xiang}\ \bibnamefont {Zhi}}, \bibinfo
  {author} {\bibfnamefont {Yu}~\bibnamefont {Liu}}, \bibinfo {author}
  {\bibfnamefont {Christian~E.}\ \bibnamefont {Matt}}, \bibinfo {author}
  {\bibfnamefont {Brinda}\ \bibnamefont {Kuthanazhi}}, \bibinfo {author}
  {\bibfnamefont {Paul~C.}\ \bibnamefont {Canfield}}, \bibinfo {author}
  {\bibfnamefont {Dmitry}\ \bibnamefont {Yarotski}}, \bibinfo {author}
  {\bibfnamefont {A.~J.}\ \bibnamefont {Taylor}}, \ and\ \bibinfo {author}
  {\bibfnamefont {Jian-Xin}\ \bibnamefont {Zhu}},\ }\bibfield  {title}
  {\enquote {\bibinfo {title} {Prediction of spin polarized fermi arcs in
  quasiparticle interference in cebi},}\ }\href {\doibase
  10.1103/PhysRevB.102.235167} {\bibfield  {journal} {\bibinfo  {journal}
  {Phys. Rev. B}\ }\textbf {\bibinfo {volume} {102}},\ \bibinfo {pages}
  {235167} (\bibinfo {year} {2020})}\BibitemShut {NoStop}%
\bibitem [{\citenamefont {Fang}\ \emph {et~al.}(2020)\citenamefont {Fang},
  \citenamefont {Tang}, \citenamefont {Ruan}, \citenamefont {Zhang},
  \citenamefont {Zhang}, \citenamefont {Gu}, \citenamefont {Zhao},
  \citenamefont {Han}, \citenamefont {Tian}, \citenamefont {Qian} \emph
  {et~al.}}]{fang2020magnetic}%
  \BibitemOpen
  \bibfield  {author} {\bibinfo {author} {\bibfnamefont {Y}~\bibnamefont
  {Fang}}, \bibinfo {author} {\bibfnamefont {F}~\bibnamefont {Tang}}, \bibinfo
  {author} {\bibfnamefont {YR}~\bibnamefont {Ruan}}, \bibinfo {author}
  {\bibfnamefont {JM}~\bibnamefont {Zhang}}, \bibinfo {author} {\bibfnamefont
  {H}~\bibnamefont {Zhang}}, \bibinfo {author} {\bibfnamefont {H}~\bibnamefont
  {Gu}}, \bibinfo {author} {\bibfnamefont {WY}~\bibnamefont {Zhao}}, \bibinfo
  {author} {\bibfnamefont {ZD}~\bibnamefont {Han}}, \bibinfo {author}
  {\bibfnamefont {W}~\bibnamefont {Tian}}, \bibinfo {author} {\bibfnamefont
  {B}~\bibnamefont {Qian}},  \emph {et~al.},\ }\bibfield  {title} {\enquote
  {\bibinfo {title} {Magnetic-field-induced nontrivial electronic state in the
  kondo-lattice semimetal cesb},}\ }\href@noop {} {\bibfield  {journal}
  {\bibinfo  {journal} {Physical Review B}\ }\textbf {\bibinfo {volume}
  {101}},\ \bibinfo {pages} {094424} (\bibinfo {year} {2020})}\BibitemShut
  {NoStop}%
\bibitem [{\citenamefont {Li}\ \emph {et~al.}(2017)\citenamefont {Li},
  \citenamefont {Xu}, \citenamefont {Ning}, \citenamefont {Su}, \citenamefont
  {Iitaka}, \citenamefont {Tohyama},\ and\ \citenamefont
  {Zhang}}]{li2017predicted}%
  \BibitemOpen
  \bibfield  {author} {\bibinfo {author} {\bibfnamefont {Zhi}\ \bibnamefont
  {Li}}, \bibinfo {author} {\bibfnamefont {Dan-Dan}\ \bibnamefont {Xu}},
  \bibinfo {author} {\bibfnamefont {Shu-Yu}\ \bibnamefont {Ning}}, \bibinfo
  {author} {\bibfnamefont {Haibin}\ \bibnamefont {Su}}, \bibinfo {author}
  {\bibfnamefont {Toshiaki}\ \bibnamefont {Iitaka}}, \bibinfo {author}
  {\bibfnamefont {Takami}\ \bibnamefont {Tohyama}}, \ and\ \bibinfo {author}
  {\bibfnamefont {Jiu-Xing}\ \bibnamefont {Zhang}},\ }\bibfield  {title}
  {\enquote {\bibinfo {title} {Predicted weyl fermions in magnetic gdbi and
  gdsb},}\ }\href@noop {} {\bibfield  {journal} {\bibinfo  {journal}
  {International Journal of Modern Physics B}\ }\textbf {\bibinfo {volume}
  {31}},\ \bibinfo {pages} {1750217} (\bibinfo {year} {2017})}\BibitemShut
  {NoStop}%
\bibitem [{\citenamefont {Matt}\ \emph {et~al.}(2022)\citenamefont {Matt},
  \citenamefont {Liu}, \citenamefont {Pirie}, \citenamefont {Drucker},
  \citenamefont {Jo}, \citenamefont {Kuthanazhi}, \citenamefont {Huang},
  \citenamefont {Lane}, \citenamefont {Zhu}, \citenamefont {Canfield} \emph
  {et~al.}}]{matt2020}%
  \BibitemOpen
  \bibfield  {author} {\bibinfo {author} {\bibfnamefont {Christian~E}\
  \bibnamefont {Matt}}, \bibinfo {author} {\bibfnamefont {Yu}~\bibnamefont
  {Liu}}, \bibinfo {author} {\bibfnamefont {Harris}\ \bibnamefont {Pirie}},
  \bibinfo {author} {\bibfnamefont {Nathan~C}\ \bibnamefont {Drucker}},
  \bibinfo {author} {\bibfnamefont {Na~Hyun}\ \bibnamefont {Jo}}, \bibinfo
  {author} {\bibfnamefont {Brinda}\ \bibnamefont {Kuthanazhi}}, \bibinfo
  {author} {\bibfnamefont {Zhao}\ \bibnamefont {Huang}}, \bibinfo {author}
  {\bibfnamefont {Christopher}\ \bibnamefont {Lane}}, \bibinfo {author}
  {\bibfnamefont {Jian-Xin}\ \bibnamefont {Zhu}}, \bibinfo {author}
  {\bibfnamefont {Paul~C}\ \bibnamefont {Canfield}},  \emph {et~al.},\
  }\bibfield  {title} {\enquote {\bibinfo {title} {Spin-polarized imaging of
  strongly interacting fermions in the ferrimagnetic state of the weyl
  candidate cebi},}\ }\href@noop {} {\bibfield  {journal} {\bibinfo  {journal}
  {Physical Review B}\ }\textbf {\bibinfo {volume} {105}},\ \bibinfo {pages}
  {085134} (\bibinfo {year} {2022})}\BibitemShut {NoStop}%
\bibitem [{\citenamefont {Manfrinetti}\ \emph {et~al.}(2009)\citenamefont
  {Manfrinetti}, \citenamefont {Provino}, \citenamefont {Morozkin},\ and\
  \citenamefont {Isnard}}]{manfrinetti2009magnetic}%
  \BibitemOpen
  \bibfield  {author} {\bibinfo {author} {\bibfnamefont {Pietro}\ \bibnamefont
  {Manfrinetti}}, \bibinfo {author} {\bibfnamefont {A}~\bibnamefont {Provino}},
  \bibinfo {author} {\bibfnamefont {AV}~\bibnamefont {Morozkin}}, \ and\
  \bibinfo {author} {\bibfnamefont {Olivier}\ \bibnamefont {Isnard}},\
  }\bibfield  {title} {\enquote {\bibinfo {title} {Magnetic structure of the
  nacl-type ndsb compound},}\ }\href@noop {} {\bibfield  {journal} {\bibinfo
  {journal} {Journal of Alloys and Compounds}\ }\textbf {\bibinfo {volume}
  {487}},\ \bibinfo {pages} {L28--L29} (\bibinfo {year} {2009})}\BibitemShut
  {NoStop}%
\bibitem [{\citenamefont {Wan}\ \emph {et~al.}(2011)\citenamefont {Wan},
  \citenamefont {Turner}, \citenamefont {Vishwanath},\ and\ \citenamefont
  {Savrasov}}]{wan2011topological}%
  \BibitemOpen
  \bibfield  {author} {\bibinfo {author} {\bibfnamefont {Xiangang}\
  \bibnamefont {Wan}}, \bibinfo {author} {\bibfnamefont {Ari~M.}\ \bibnamefont
  {Turner}}, \bibinfo {author} {\bibfnamefont {Ashvin}\ \bibnamefont
  {Vishwanath}}, \ and\ \bibinfo {author} {\bibfnamefont {Sergey~Y.}\
  \bibnamefont {Savrasov}},\ }\bibfield  {title} {\enquote {\bibinfo {title}
  {Topological semimetal and fermi-arc surface states in the electronic
  structure of pyrochlore iridates},}\ }\href {\doibase
  10.1103/PhysRevB.83.205101} {\bibfield  {journal} {\bibinfo  {journal} {Phys.
  Rev. B}\ }\textbf {\bibinfo {volume} {83}},\ \bibinfo {pages} {205101}
  (\bibinfo {year} {2011})}\BibitemShut {NoStop}%
\bibitem [{\citenamefont {Schrunk}\ \emph {et~al.}(2022)\citenamefont
  {Schrunk}, \citenamefont {Kushnirenko}, \citenamefont {Kuthanazhi},
  \citenamefont {Ahn}, \citenamefont {Wang}, \citenamefont {O’Leary},
  \citenamefont {Lee}, \citenamefont {Eaton}, \citenamefont {Fedorov},
  \citenamefont {Lou}, \citenamefont {Voroshnin}, \citenamefont {Clark},
  \citenamefont {Sánchez-Barriga}, \citenamefont {Bud’ko}, \citenamefont
  {Slager}, \citenamefont {Canfield},\ and\ \citenamefont
  {Kaminski}}]{SchrunkNature2022}%
  \BibitemOpen
  \bibfield  {author} {\bibinfo {author} {\bibfnamefont {Benjamin}\
  \bibnamefont {Schrunk}}, \bibinfo {author} {\bibfnamefont {Yevhen}\
  \bibnamefont {Kushnirenko}}, \bibinfo {author} {\bibfnamefont {Brinda}\
  \bibnamefont {Kuthanazhi}}, \bibinfo {author} {\bibfnamefont {Junyeong}\
  \bibnamefont {Ahn}}, \bibinfo {author} {\bibfnamefont {Lin-Lin}\ \bibnamefont
  {Wang}}, \bibinfo {author} {\bibfnamefont {Evan}\ \bibnamefont {O’Leary}},
  \bibinfo {author} {\bibfnamefont {Kyungchan}\ \bibnamefont {Lee}}, \bibinfo
  {author} {\bibfnamefont {Andrew}\ \bibnamefont {Eaton}}, \bibinfo {author}
  {\bibfnamefont {Alexander}\ \bibnamefont {Fedorov}}, \bibinfo {author}
  {\bibfnamefont {Rui}\ \bibnamefont {Lou}}, \bibinfo {author} {\bibfnamefont
  {Vladimir}\ \bibnamefont {Voroshnin}}, \bibinfo {author} {\bibfnamefont
  {Oliver~J.}\ \bibnamefont {Clark}}, \bibinfo {author} {\bibfnamefont {Jamie}\
  \bibnamefont {Sánchez-Barriga}}, \bibinfo {author} {\bibfnamefont
  {Sergey~L.}\ \bibnamefont {Bud’ko}}, \bibinfo {author} {\bibfnamefont
  {Robert-Jan}\ \bibnamefont {Slager}}, \bibinfo {author} {\bibfnamefont
  {Paul~C.}\ \bibnamefont {Canfield}}, \ and\ \bibinfo {author} {\bibfnamefont
  {Adam}\ \bibnamefont {Kaminski}},\ }\bibfield  {title} {\enquote {\bibinfo
  {title} {Emergence of fermi arcs due to magnetic splitting in an
  antiferromagnet},}\ }\href@noop {} {\bibfield  {journal} {\bibinfo  {journal}
  {Nature}\ }\textbf {\bibinfo {volume} {603}},\ \bibinfo {pages} {610–615}
  (\bibinfo {year} {2022})}\BibitemShut {NoStop}%
\bibitem [{\citenamefont {Wang}\ \emph {et~al.}(2022)\citenamefont {Wang},
  \citenamefont {Ahn}, \citenamefont {Slager}, \citenamefont {Kushnirenko},
  \citenamefont {Ueland}, \citenamefont {Sapkota}, \citenamefont {Schrunk},
  \citenamefont {Kuthanazhi}, \citenamefont {McQueeney}, \citenamefont
  {Canfield} \emph {et~al.}}]{wang2022multi}%
  \BibitemOpen
  \bibfield  {author} {\bibinfo {author} {\bibfnamefont {L-L}\ \bibnamefont
  {Wang}}, \bibinfo {author} {\bibfnamefont {J}~\bibnamefont {Ahn}}, \bibinfo
  {author} {\bibfnamefont {R-J}\ \bibnamefont {Slager}}, \bibinfo {author}
  {\bibfnamefont {Y}~\bibnamefont {Kushnirenko}}, \bibinfo {author}
  {\bibfnamefont {BG}~\bibnamefont {Ueland}}, \bibinfo {author} {\bibfnamefont
  {A}~\bibnamefont {Sapkota}}, \bibinfo {author} {\bibfnamefont
  {B}~\bibnamefont {Schrunk}}, \bibinfo {author} {\bibfnamefont
  {B}~\bibnamefont {Kuthanazhi}}, \bibinfo {author} {\bibfnamefont
  {RJ}~\bibnamefont {McQueeney}}, \bibinfo {author} {\bibfnamefont
  {PC}~\bibnamefont {Canfield}},  \emph {et~al.},\ }\bibfield  {title}
  {\enquote {\bibinfo {title} {Multi-q origin of unconventional surface states
  in a high-symmetry lattice},}\ }\href@noop {} {\bibfield  {journal} {\bibinfo
   {journal} {arXiv preprint arXiv:2203.12541}\ } (\bibinfo {year}
  {2022})}\BibitemShut {NoStop}%
\bibitem [{\citenamefont {Kuthanazhi}\ \emph {et~al.}(2019)\citenamefont
  {Kuthanazhi}, \citenamefont {Jo}, \citenamefont {Xiang}, \citenamefont
  {Bud'ko},\ and\ \citenamefont {Canfield}}]{kuthanazhi2019metamagnetism}%
  \BibitemOpen
  \bibfield  {author} {\bibinfo {author} {\bibfnamefont {Brinda}\ \bibnamefont
  {Kuthanazhi}}, \bibinfo {author} {\bibfnamefont {Na~Hyun}\ \bibnamefont
  {Jo}}, \bibinfo {author} {\bibfnamefont {Li}~\bibnamefont {Xiang}}, \bibinfo
  {author} {\bibfnamefont {Sergey~L}\ \bibnamefont {Bud'ko}}, \ and\ \bibinfo
  {author} {\bibfnamefont {Paul~C}\ \bibnamefont {Canfield}},\ }\bibfield
  {title} {\enquote {\bibinfo {title} {Metamagnetism and magnetoresistance in
  cebi single crystals},}\ }\href@noop {} {\bibfield  {journal} {\bibinfo
  {journal} {arXiv preprint arXiv:1912.08896}\ } (\bibinfo {year}
  {2019})}\BibitemShut {NoStop}%
\bibitem [{\citenamefont {Sakhya}\ \emph {et~al.}(2021)\citenamefont {Sakhya},
  \citenamefont {Paulose}, \citenamefont {Thamizhavel},\ and\ \citenamefont
  {Maiti}}]{sakhya2021evidence}%
  \BibitemOpen
  \bibfield  {author} {\bibinfo {author} {\bibfnamefont {Anup~Pradhan}\
  \bibnamefont {Sakhya}}, \bibinfo {author} {\bibfnamefont {PL}~\bibnamefont
  {Paulose}}, \bibinfo {author} {\bibfnamefont {A}~\bibnamefont {Thamizhavel}},
  \ and\ \bibinfo {author} {\bibfnamefont {Kalobaran}\ \bibnamefont {Maiti}},\
  }\bibfield  {title} {\enquote {\bibinfo {title} {Evidence of nontrivial berry
  phase and kondo physics in smbi},}\ }\href@noop {} {\bibfield  {journal}
  {\bibinfo  {journal} {Physical Review Materials}\ }\textbf {\bibinfo {volume}
  {5}},\ \bibinfo {pages} {054201} (\bibinfo {year} {2021})}\BibitemShut
  {NoStop}%
\bibitem [{\citenamefont {Canfield}\ \emph {et~al.}(2016)\citenamefont
  {Canfield}, \citenamefont {Kong}, \citenamefont {Kaluarachchi},\ and\
  \citenamefont {Jo}}]{Canfield2016Use}%
  \BibitemOpen
  \bibfield  {author} {\bibinfo {author} {\bibfnamefont {P.~C.}\ \bibnamefont
  {Canfield}}, \bibinfo {author} {\bibfnamefont {Tai}\ \bibnamefont {Kong}},
  \bibinfo {author} {\bibfnamefont {Udhara~S}\ \bibnamefont {Kaluarachchi}}, \
  and\ \bibinfo {author} {\bibfnamefont {Na~Hyun}\ \bibnamefont {Jo}},\
  }\bibfield  {title} {\enquote {\bibinfo {title} {{Use of frit-disc crucibles
  for routine and exploratory solution growth of single crystalline
  samples}},}\ }\href@noop {} {\bibfield  {journal} {\bibinfo  {journal}
  {Philosophical Magazine}\ }\textbf {\bibinfo {volume} {96}},\ \bibinfo
  {pages} {84--92} (\bibinfo {year} {2016})}\BibitemShut {NoStop}%
\bibitem [{\citenamefont {Canfield}(2019)}]{Canfield_2019}%
  \BibitemOpen
  \bibfield  {author} {\bibinfo {author} {\bibfnamefont {Paul~C}\ \bibnamefont
  {Canfield}},\ }\bibfield  {title} {\enquote {\bibinfo {title} {New materials
  physics},}\ }\href {\doibase 10.1088/1361-6633/ab514b} {\bibfield  {journal}
  {\bibinfo  {journal} {Reports on Progress in Physics}\ }\textbf {\bibinfo
  {volume} {83}},\ \bibinfo {pages} {016501} (\bibinfo {year}
  {2019})}\BibitemShut {NoStop}%
\bibitem [{\citenamefont {Jiang}\ \emph {et~al.}(2014)\citenamefont {Jiang},
  \citenamefont {Mou}, \citenamefont {Wu}, \citenamefont {Huang}, \citenamefont
  {McMillen}, \citenamefont {Kolis}, \citenamefont {Giesber}, \citenamefont
  {Egan},\ and\ \citenamefont {Kaminski}}]{jiang2014tunable}%
  \BibitemOpen
  \bibfield  {author} {\bibinfo {author} {\bibfnamefont {Rui}\ \bibnamefont
  {Jiang}}, \bibinfo {author} {\bibfnamefont {Daixiang}\ \bibnamefont {Mou}},
  \bibinfo {author} {\bibfnamefont {Yun}\ \bibnamefont {Wu}}, \bibinfo {author}
  {\bibfnamefont {Lunan}\ \bibnamefont {Huang}}, \bibinfo {author}
  {\bibfnamefont {Colin~D.}\ \bibnamefont {McMillen}}, \bibinfo {author}
  {\bibfnamefont {Joseph}\ \bibnamefont {Kolis}}, \bibinfo {author}
  {\bibfnamefont {Henry~G.}\ \bibnamefont {Giesber}}, \bibinfo {author}
  {\bibfnamefont {John~J.}\ \bibnamefont {Egan}}, \ and\ \bibinfo {author}
  {\bibfnamefont {Adam}\ \bibnamefont {Kaminski}},\ }\bibfield  {title}
  {\enquote {\bibinfo {title} {Tunable vacuum ultraviolet laser based
  spectrometer for angle resolved photoemission spectroscopy},}\ }\href
  {\doibase http://dx.doi.org/10.1063/1.4867517} {\bibfield  {journal}
  {\bibinfo  {journal} {Review of Scientific Instruments}\ }\textbf {\bibinfo
  {volume} {85}},\ \bibinfo {eid} {033902} (\bibinfo {year}
  {2014})}\BibitemShut {NoStop}%
\bibitem [{\citenamefont {Hohenberg}\ and\ \citenamefont
  {Kohn}(1964)}]{hohenberg1964inhomogeneous}%
  \BibitemOpen
  \bibfield  {author} {\bibinfo {author} {\bibfnamefont {P.}~\bibnamefont
  {Hohenberg}}\ and\ \bibinfo {author} {\bibfnamefont {W.}~\bibnamefont
  {Kohn}},\ }\bibfield  {title} {\enquote {\bibinfo {title} {Inhomogeneous
  electron gas},}\ }\href {\doibase 10.1103/PhysRev.136.B864} {\bibfield
  {journal} {\bibinfo  {journal} {Phys. Rev.}\ }\textbf {\bibinfo {volume}
  {136}},\ \bibinfo {pages} {B864--B871} (\bibinfo {year} {1964})}\BibitemShut
  {NoStop}%
\bibitem [{\citenamefont {Kohn}\ and\ \citenamefont
  {Sham}(1965)}]{kohn1965self}%
  \BibitemOpen
  \bibfield  {author} {\bibinfo {author} {\bibfnamefont {W.}~\bibnamefont
  {Kohn}}\ and\ \bibinfo {author} {\bibfnamefont {L.~J.}\ \bibnamefont
  {Sham}},\ }\bibfield  {title} {\enquote {\bibinfo {title} {Self-consistent
  equations including exchange and correlation effects},}\ }\href {\doibase
  10.1103/PhysRev.140.A1133} {\bibfield  {journal} {\bibinfo  {journal} {Phys.
  Rev.}\ }\textbf {\bibinfo {volume} {140}},\ \bibinfo {pages} {A1133--A1138}
  (\bibinfo {year} {1965})}\BibitemShut {NoStop}%
\bibitem [{\citenamefont {Perdew}\ \emph {et~al.}(1996)\citenamefont {Perdew},
  \citenamefont {Burke},\ and\ \citenamefont
  {Ernzerhof}}]{Perdew1996Generalized}%
  \BibitemOpen
  \bibfield  {author} {\bibinfo {author} {\bibfnamefont {John~P.}\ \bibnamefont
  {Perdew}}, \bibinfo {author} {\bibfnamefont {Kieron}\ \bibnamefont {Burke}},
  \ and\ \bibinfo {author} {\bibfnamefont {Matthias}\ \bibnamefont
  {Ernzerhof}},\ }\bibfield  {title} {\enquote {\bibinfo {title} {Generalized
  gradient approximation made simple},}\ }\href {\doibase
  10.1103/PhysRevLett.77.3865} {\bibfield  {journal} {\bibinfo  {journal}
  {Phys. Rev. Lett.}\ }\textbf {\bibinfo {volume} {77}},\ \bibinfo {pages}
  {3865--3868} (\bibinfo {year} {1996})}\BibitemShut {NoStop}%
\bibitem [{\citenamefont {Bl\"ochl}(1994)}]{Blochl1994Projector}%
  \BibitemOpen
  \bibfield  {author} {\bibinfo {author} {\bibfnamefont {P.~E.}\ \bibnamefont
  {Bl\"ochl}},\ }\bibfield  {title} {\enquote {\bibinfo {title} {Projector
  augmented-wave method},}\ }\href {\doibase 10.1103/PhysRevB.50.17953}
  {\bibfield  {journal} {\bibinfo  {journal} {Phys. Rev. B}\ }\textbf {\bibinfo
  {volume} {50}},\ \bibinfo {pages} {17953--17979} (\bibinfo {year}
  {1994})}\BibitemShut {NoStop}%
\bibitem [{\citenamefont {Kresse}\ and\ \citenamefont
  {Furthmüller}(1996)}]{Kresse1996Efficiency}%
  \BibitemOpen
  \bibfield  {author} {\bibinfo {author} {\bibfnamefont {G.}~\bibnamefont
  {Kresse}}\ and\ \bibinfo {author} {\bibfnamefont {J.}~\bibnamefont
  {Furthmüller}},\ }\bibfield  {title} {\enquote {\bibinfo {title} {Efficiency
  of ab-initio total energy calculations for metals and semiconductors using a
  plane-wave basis set},}\ }\href {\doibase
  http://dx.doi.org/10.1016/0927-0256(96)00008-0} {\bibfield  {journal}
  {\bibinfo  {journal} {Computational Materials Science}\ }\textbf {\bibinfo
  {volume} {6}},\ \bibinfo {pages} {15 -- 50} (\bibinfo {year}
  {1996})}\BibitemShut {NoStop}%
\bibitem [{\citenamefont {Kresse}\ and\ \citenamefont
  {Furthm\"uller}(1996)}]{Kresse1996Efficient}%
  \BibitemOpen
  \bibfield  {author} {\bibinfo {author} {\bibfnamefont {G.}~\bibnamefont
  {Kresse}}\ and\ \bibinfo {author} {\bibfnamefont {J.}~\bibnamefont
  {Furthm\"uller}},\ }\bibfield  {title} {\enquote {\bibinfo {title} {Efficient
  iterative schemes for \textit{ab initio} total-energy calculations using a
  plane-wave basis set},}\ }\href {\doibase 10.1103/PhysRevB.54.11169}
  {\bibfield  {journal} {\bibinfo  {journal} {Phys. Rev. B}\ }\textbf {\bibinfo
  {volume} {54}},\ \bibinfo {pages} {11169--11186} (\bibinfo {year}
  {1996})}\BibitemShut {NoStop}%
\bibitem [{\citenamefont {Marzari}\ and\ \citenamefont
  {Vanderbilt}(1997)}]{Marzari1997Maximally}%
  \BibitemOpen
  \bibfield  {author} {\bibinfo {author} {\bibfnamefont {Nicola}\ \bibnamefont
  {Marzari}}\ and\ \bibinfo {author} {\bibfnamefont {David}\ \bibnamefont
  {Vanderbilt}},\ }\bibfield  {title} {\enquote {\bibinfo {title} {Maximally
  localized generalized wannier functions for composite energy bands},}\ }\href
  {\doibase 10.1103/PhysRevB.56.12847} {\bibfield  {journal} {\bibinfo
  {journal} {Phys. Rev. B}\ }\textbf {\bibinfo {volume} {56}},\ \bibinfo
  {pages} {12847--12865} (\bibinfo {year} {1997})}\BibitemShut {NoStop}%
\bibitem [{\citenamefont {Souza}\ \emph {et~al.}(2001)\citenamefont {Souza},
  \citenamefont {Marzari},\ and\ \citenamefont
  {Vanderbilt}}]{Souza2001Maximally}%
  \BibitemOpen
  \bibfield  {author} {\bibinfo {author} {\bibfnamefont {Ivo}\ \bibnamefont
  {Souza}}, \bibinfo {author} {\bibfnamefont {Nicola}\ \bibnamefont {Marzari}},
  \ and\ \bibinfo {author} {\bibfnamefont {David}\ \bibnamefont {Vanderbilt}},\
  }\bibfield  {title} {\enquote {\bibinfo {title} {Maximally localized wannier
  functions for entangled energy bands},}\ }\href {\doibase
  10.1103/PhysRevB.65.035109} {\bibfield  {journal} {\bibinfo  {journal} {Phys.
  Rev. B}\ }\textbf {\bibinfo {volume} {65}},\ \bibinfo {pages} {035109}
  (\bibinfo {year} {2001})}\BibitemShut {NoStop}%
\bibitem [{\citenamefont {Sancho}\ \emph {et~al.}(1984)\citenamefont {Sancho},
  \citenamefont {Sancho},\ and\ \citenamefont {Rubio}}]{Sancho1984Quick}%
  \BibitemOpen
  \bibfield  {author} {\bibinfo {author} {\bibfnamefont {M~P~Lopez}\
  \bibnamefont {Sancho}}, \bibinfo {author} {\bibfnamefont {J~M~Lopez}\
  \bibnamefont {Sancho}}, \ and\ \bibinfo {author} {\bibfnamefont
  {J}~\bibnamefont {Rubio}},\ }\bibfield  {title} {\enquote {\bibinfo {title}
  {Quick iterative scheme for the calculation of transfer matrices: application
  to mo (100)},}\ }\href {http://stacks.iop.org/0305-4608/14/i=5/a=016}
  {\bibfield  {journal} {\bibinfo  {journal} {Journal of Physics F: Metal
  Physics}\ }\textbf {\bibinfo {volume} {14}},\ \bibinfo {pages} {1205}
  (\bibinfo {year} {1984})}\BibitemShut {NoStop}%
\bibitem [{\citenamefont {Sancho}\ \emph {et~al.}(1985)\citenamefont {Sancho},
  \citenamefont {Sancho}, \citenamefont {Sancho},\ and\ \citenamefont
  {Rubio}}]{Sancho1985Highly}%
  \BibitemOpen
  \bibfield  {author} {\bibinfo {author} {\bibfnamefont {M~P~Lopez}\
  \bibnamefont {Sancho}}, \bibinfo {author} {\bibfnamefont {J~M~Lopez}\
  \bibnamefont {Sancho}}, \bibinfo {author} {\bibfnamefont {J~M~L}\
  \bibnamefont {Sancho}}, \ and\ \bibinfo {author} {\bibfnamefont
  {J}~\bibnamefont {Rubio}},\ }\bibfield  {title} {\enquote {\bibinfo {title}
  {Highly convergent schemes for the calculation of bulk and surface green
  functions},}\ }\href {http://stacks.iop.org/0305-4608/15/i=4/a=009}
  {\bibfield  {journal} {\bibinfo  {journal} {Journal of Physics F: Metal
  Physics}\ }\textbf {\bibinfo {volume} {15}},\ \bibinfo {pages} {851}
  (\bibinfo {year} {1985})}\BibitemShut {NoStop}%
\bibitem [{\citenamefont {Wu}\ \emph {et~al.}(2018)\citenamefont {Wu},
  \citenamefont {Zhang}, \citenamefont {Song}, \citenamefont {Troyer},\ and\
  \citenamefont {Soluyanov}}]{wu2018wanniertools}%
  \BibitemOpen
  \bibfield  {author} {\bibinfo {author} {\bibfnamefont {QuanSheng}\
  \bibnamefont {Wu}}, \bibinfo {author} {\bibfnamefont {ShengNan}\ \bibnamefont
  {Zhang}}, \bibinfo {author} {\bibfnamefont {Hai-Feng}\ \bibnamefont {Song}},
  \bibinfo {author} {\bibfnamefont {Matthias}\ \bibnamefont {Troyer}}, \ and\
  \bibinfo {author} {\bibfnamefont {Alexey~A}\ \bibnamefont {Soluyanov}},\
  }\bibfield  {title} {\enquote {\bibinfo {title} {Wanniertools: An open-source
  software package for novel topological materials},}\ }\href@noop {}
  {\bibfield  {journal} {\bibinfo  {journal} {Computer Physics Communications}\
  }\textbf {\bibinfo {volume} {224}},\ \bibinfo {pages} {405--416} (\bibinfo
  {year} {2018})}\BibitemShut {NoStop}%
\bibitem [{\citenamefont {Monkhorst}\ and\ \citenamefont
  {Pack}(1976)}]{Monkhorst1976Special}%
  \BibitemOpen
  \bibfield  {author} {\bibinfo {author} {\bibfnamefont {Hendrik~J.}\
  \bibnamefont {Monkhorst}}\ and\ \bibinfo {author} {\bibfnamefont {James~D.}\
  \bibnamefont {Pack}},\ }\bibfield  {title} {\enquote {\bibinfo {title}
  {Special points for brillouin-zone integrations},}\ }\href {\doibase
  10.1103/PhysRevB.13.5188} {\bibfield  {journal} {\bibinfo  {journal} {Phys.
  Rev. B}\ }\textbf {\bibinfo {volume} {13}},\ \bibinfo {pages} {5188--5192}
  (\bibinfo {year} {1976})}\BibitemShut {NoStop}%
\bibitem [{\citenamefont {Dudarev}\ \emph {et~al.}(1998)\citenamefont
  {Dudarev}, \citenamefont {Botton}, \citenamefont {Savrasov}, \citenamefont
  {Humphreys},\ and\ \citenamefont {Sutton}}]{dudarev1998electron}%
  \BibitemOpen
  \bibfield  {author} {\bibinfo {author} {\bibfnamefont {Sergei~L}\
  \bibnamefont {Dudarev}}, \bibinfo {author} {\bibfnamefont {Gianluigi~A}\
  \bibnamefont {Botton}}, \bibinfo {author} {\bibfnamefont {Sergey~Y}\
  \bibnamefont {Savrasov}}, \bibinfo {author} {\bibfnamefont {CJ}~\bibnamefont
  {Humphreys}}, \ and\ \bibinfo {author} {\bibfnamefont {Adrian~P}\
  \bibnamefont {Sutton}},\ }\bibfield  {title} {\enquote {\bibinfo {title}
  {Electron-energy-loss spectra and the structural stability of nickel oxide:
  An lsda+ u study},}\ }\href@noop {} {\bibfield  {journal} {\bibinfo
  {journal} {Physical Review B}\ }\textbf {\bibinfo {volume} {57}},\ \bibinfo
  {pages} {1505} (\bibinfo {year} {1998})}\BibitemShut {NoStop}%
\end{thebibliography}%

\section*{Acknowledgments}
ARPES measurements were supported by the U.S. Department of Energy, Office of Basic Energy Sciences, Division of Materials Science and Engineering. Ames Laboratory is operated for the U.S. Department of Energy by Iowa State University under Contract No. DE-AC02-07CH11358. Crystal growth, characterization and DFT calculations were supported by Center for the Advancement of Topological Semimetals (CATS), an Energy Frontier Research Center funded by the U.S. DOE, Office of Basic Energy Sciences.  R.-J.~S.~acknowledges funding from Trinity College at the University of Cambridge. J.A. was supported by CATS.

\section*{Methods}
{\bf Sample growth} Single crystals of CeBi, NdBi, and SmBi were grown out of In flux. The elements with an initial stoichiometry of R$_4$Bi$_4$In$_{96}$ (R = Ce, Nd, Sm) were put into a fritted alumina crucible\cite{Canfield2016Use} and sealed in fused silica tube under partial pressure of argon. NdSb crystals were grown out of Sn flux using an initial concentration of Nd$_4$Sb$_4$Sn$_{96}$. The prepared ampules were heated up to 1100$^\circ$~C over 4 hours and held there for 5 hours. This was followed by a slow cooling to the decanting temperature over 100 hours and decanting of the excess flux using a centrifuge.\cite{Canfield_2019}. The decanting temperatures were 850$^\circ$~C for CeBi, 700 C for NdBi, and 800$^\circ$~C for both SmBi and NdSb. The cubic crystals obtained were stored and handled in a glovebox under Nitrogen atmosphere. 

{\bf ARPES measurements} ARPES data was collected using vacuum ultraviolet (VUV) laser ARPES spectrometer that consists of a Scienta DA30 electron analyzer, picosecond Ti:Sapphire oscillator and fourth-harmonic generator \cite{jiang2014tunable}. Data from the laser based ARPES were collected with 6.7~eV photon energy. Angular resolution was set at $\sim$ 0.1$^{\circ}$ and 1$^{\circ}$, along and perpendicular to the direction of the analyser slit respectively, and the energy resolution was set at 2 meV. The VUV laser beam was set to vertical polarization. The diameter of the photon beam on the sample was $\sim 20\,\mu$m. Samples were cleaved \textit{in-situ} along (001) plane at a base pressure lower than 2$\times$10$^{-11}$ Torr usually producing very flat, mirror-like surfaces. Results were reproduced using several different single crystals of each material, and extensive temperature cycling. 

{\bf DFT calculations.} All density functional theory \cite{hohenberg1964inhomogeneous, kohn1965self} (DFT) calculations with spin-orbit coupling (SOC) were performed with the PBE\cite{Perdew1996Generalized} exchange-correlation functional using a plane-wave basis set and projector augmented wave method\cite{Blochl1994Projector}, as implemented in the Vienna Ab-initio Simulation Package (VASP)\cite{Kresse1996Efficiency, Kresse1996Efficient}. Using maximally localized Wannier functions\cite{Marzari1997Maximally, Souza2001Maximally}, tight-binding models were constructed to reproduce closely the band structure including SOC within $E_F\pm$1 eV with Nd $s$-$d$-$f$ and Sb $p$ orbitals. The surface spectral function and 2D Fermi surface were calculated with the surface Green's function methods\cite{Sancho1984Quick, Sancho1985Highly} as implemented in WannierTools\cite{wu2018wanniertools}. In the DFT calculations, we used a kinetic energy cutoff of 254 eV, $\Gamma$-centered Monkhorst-Pack\cite{Monkhorst1976Special} (8x8x8) $k$-point mesh, and a Gaussian smearing of 0.05 eV. For band structure calculations, we have used the experimental lattice parameters of 6.336 \AA. To account for the strongly localized Nd 4$f$ orbitals in NdSb, an onsite Hubbard-like\cite{dudarev1998electron} $U$=6.3 eV and $J$=0.7 eV have been used. Our DFT+U+SOC calculation on NdSb 2q gives a spin moment of 2.7 $\mu_B$ and an orbital moment of 5.7 $\mu_B$ in the opposite direction, resulting in a total magnetic moment of 3.0 $\mu_B$ on Nd.

\section*{Author contributions}
Y.~K., B.~S., P.~C.~C. and A.~K. conceived and designed the experiment. Y.~K. and A.~K. performed analysis of ARPES data. J.~A., L.-L.~W. and R.-J.~S provided theoretical analysis and interpretation. L.-L.~W. performed DFT calculations. B.~K., S.~L.~B. and P.~C.~C. grew and characterized the samples. Y.~K., B.~S., E.~O., A.~E. and A.~K. performed ARPES measurements and support. The manuscript was drafted by Y.~K., A.~K., P.~C.~C.  and R.-J.~S. All authors discussed and commented on the manuscript.



\clearpage

\begin{figure*}[tb]
	\includegraphics[width=6.5 in]{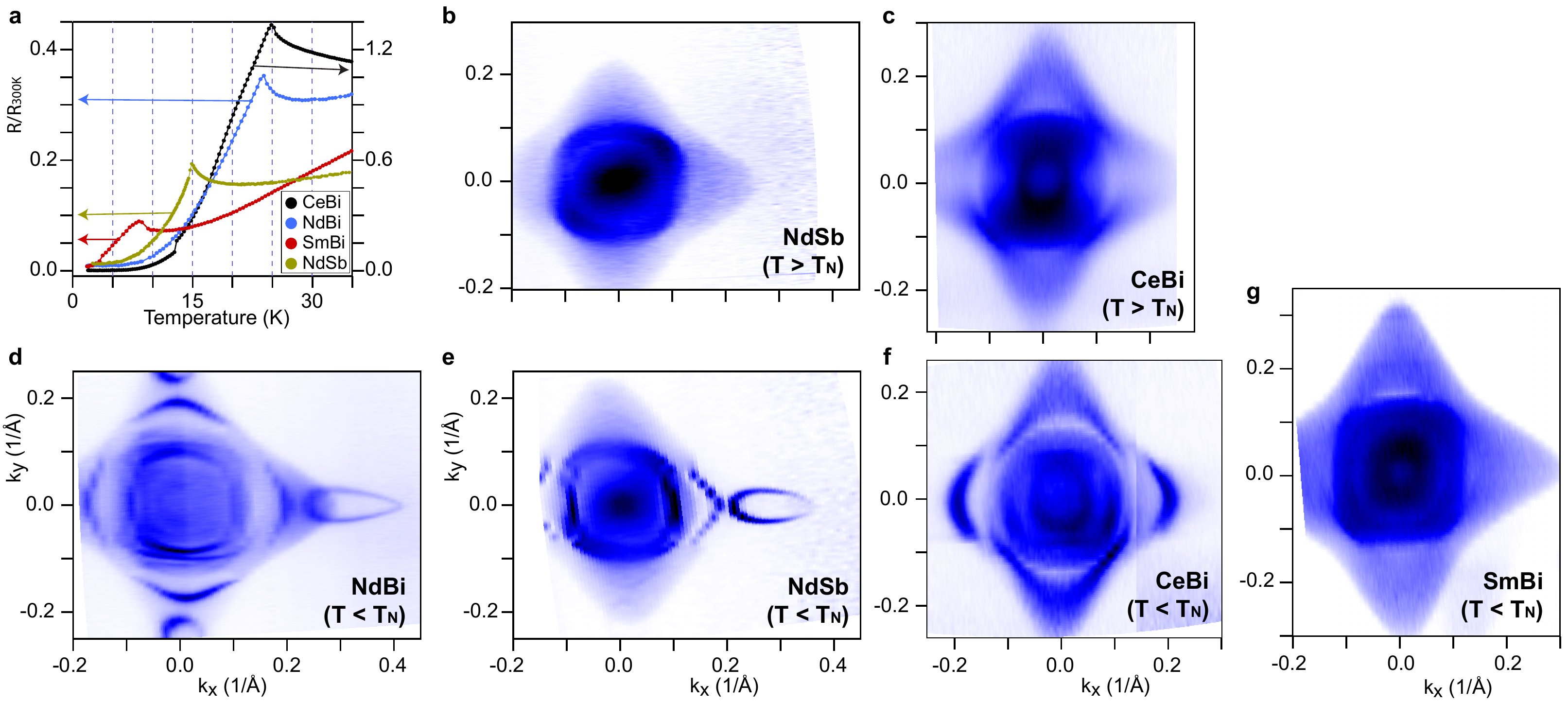}%
	\caption{{\bf $|$ Resistivity data and Fermi surface maps in pamagnetic and AFM state.}
	{\bf a}, Temperature dependence of resistivity for CeBi, NdBi, Sm Bi and NdSb normalized by value at 300 K.
    {\bf b}, Fermi surface maps of NdBi measured at T = 16K, in the paramagnetic state.
    {\bf c}, Fermi surface maps of CeBi measured at T = 30K, in the paramagnetic state.
    {\bf d} and {\bf e}, Fermi surface maps of NdBi and NdSb respectively measured in the AFM state (T = 6K). 
    {\bf f}, Fermi surface maps of CeBi in the AFM state (T = 13.5K).
    {\bf g}, Fermi surface maps of SmBi in the AFM state (T = 5K).
	\label{fig:fig1}}
\end{figure*}

\begin{figure*}[tb]
	\includegraphics[width=4.5 in]{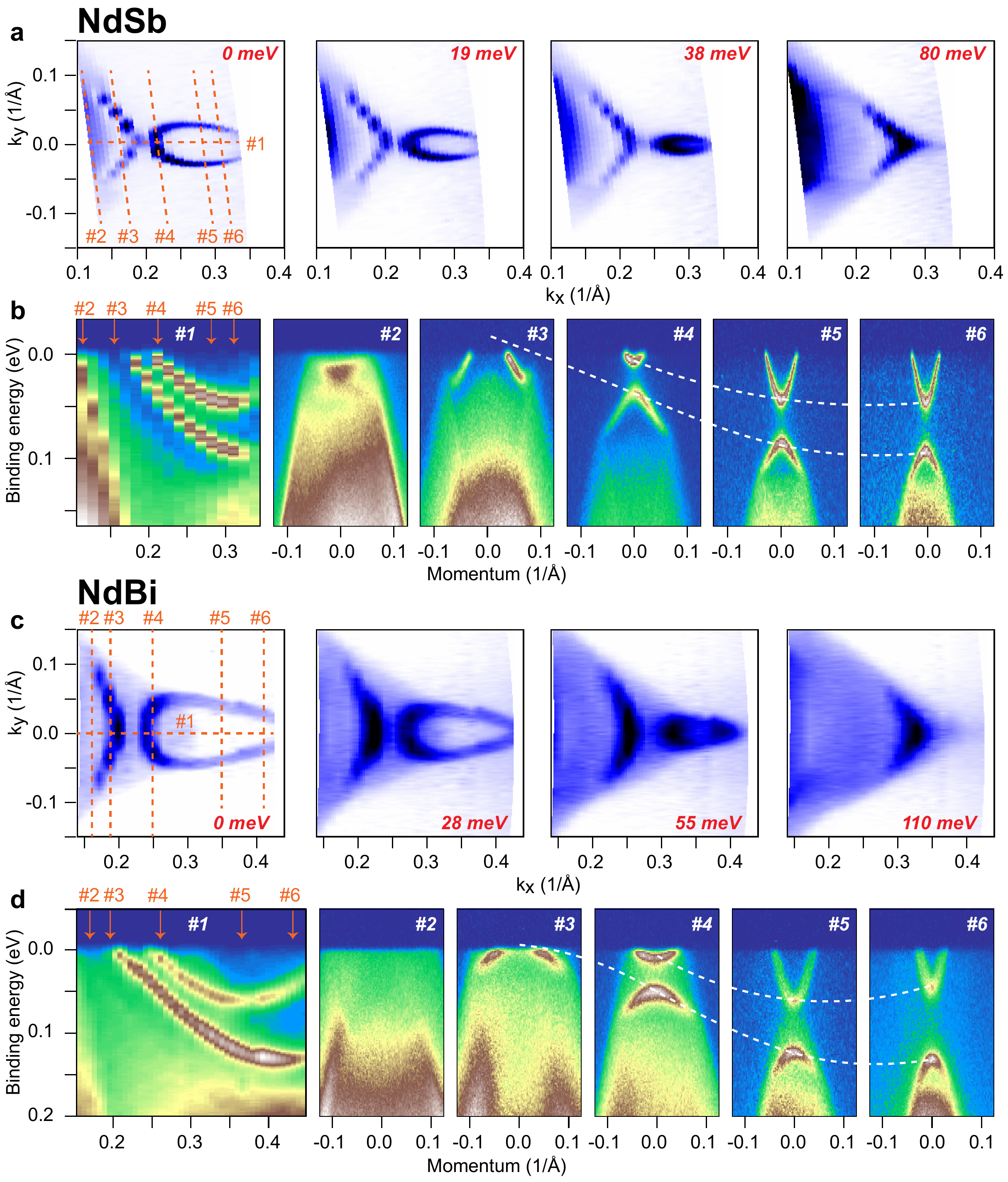}%
	\caption{{\bf $|$ Dispersion of the surface states in NdSb and NdBi in the AFM state.} 
    {\bf a}, Constant energy map for NdSb at binding energies from 0 to 80 meV measured at T = 6 K. 
    {\bf b}, Band dispersion along the cuts marked with the dashed line in ({\bf a}). The locations of cuts \#2-6 are also marked with the arrows in cut \#1.
    {\bf c}, Constant energy map for NdBi at binding energies from 0 to 110 meV measured at T = 6 K.
    {\bf d}, Band dispersion along the cuts marked with the dashed line in ({\bf c}). The locations of cuts \#2-6 are also marked with the arrows in cut \#1.
    The white dashed lines in ({\bf b} and {\bf d}) are a guide to the eye of location of top and bottom of the hole- and electron-like surface bands and reflect dispersion of the surface states seen in the corresponding cuts \#1.
	\label{fig:fig2}}
\end{figure*}

\begin{figure*}[tb]
	\includegraphics[width=6 in]{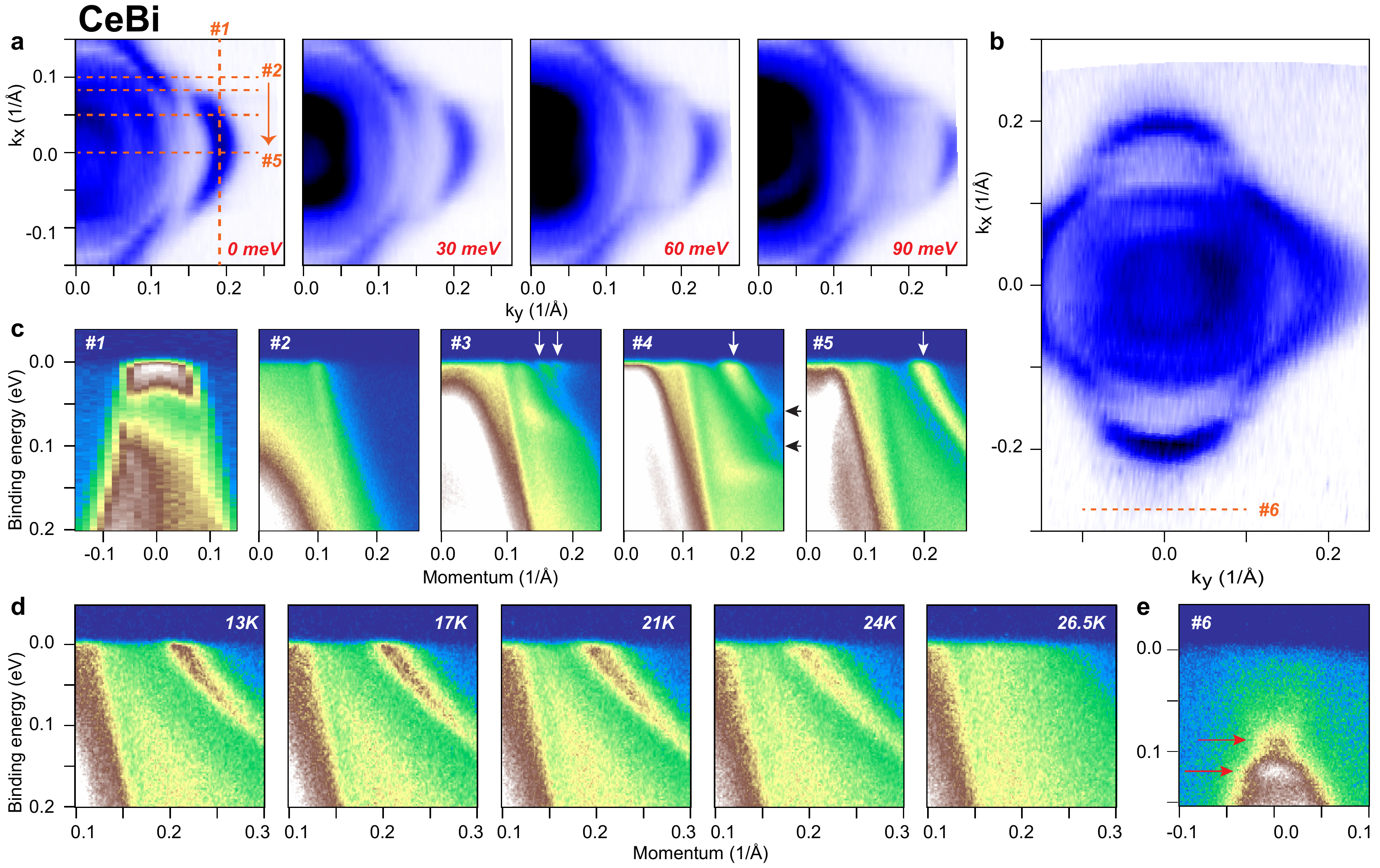}%
	\caption{{\bf $|$ Dispersion of the surface states in CeBi in the first magnetically ordered state (AFM1) at T=13.5~K.}
	{\bf a}, Constant energy plots at binding energies from 0 to 90 meV.
	{\bf b}, Fermi surface map for a domain with different orientation of magnetic ordering.
	{\bf c}, Band dispersion along the directions marked with the dashed lines in {\bf a}. Here, white arrows indicate the location of the surface state dispersions.
	{\bf d}, Temperature dependence of the band dispersion along cut \#5 in {\bf a}.
    {\bf e}, Band dispersion along the directions marked with the dashed line in {\bf b}.
	\label{fig:fig3}}
\end{figure*}

\begin{figure*}[tb]
	\includegraphics[width=5 in]{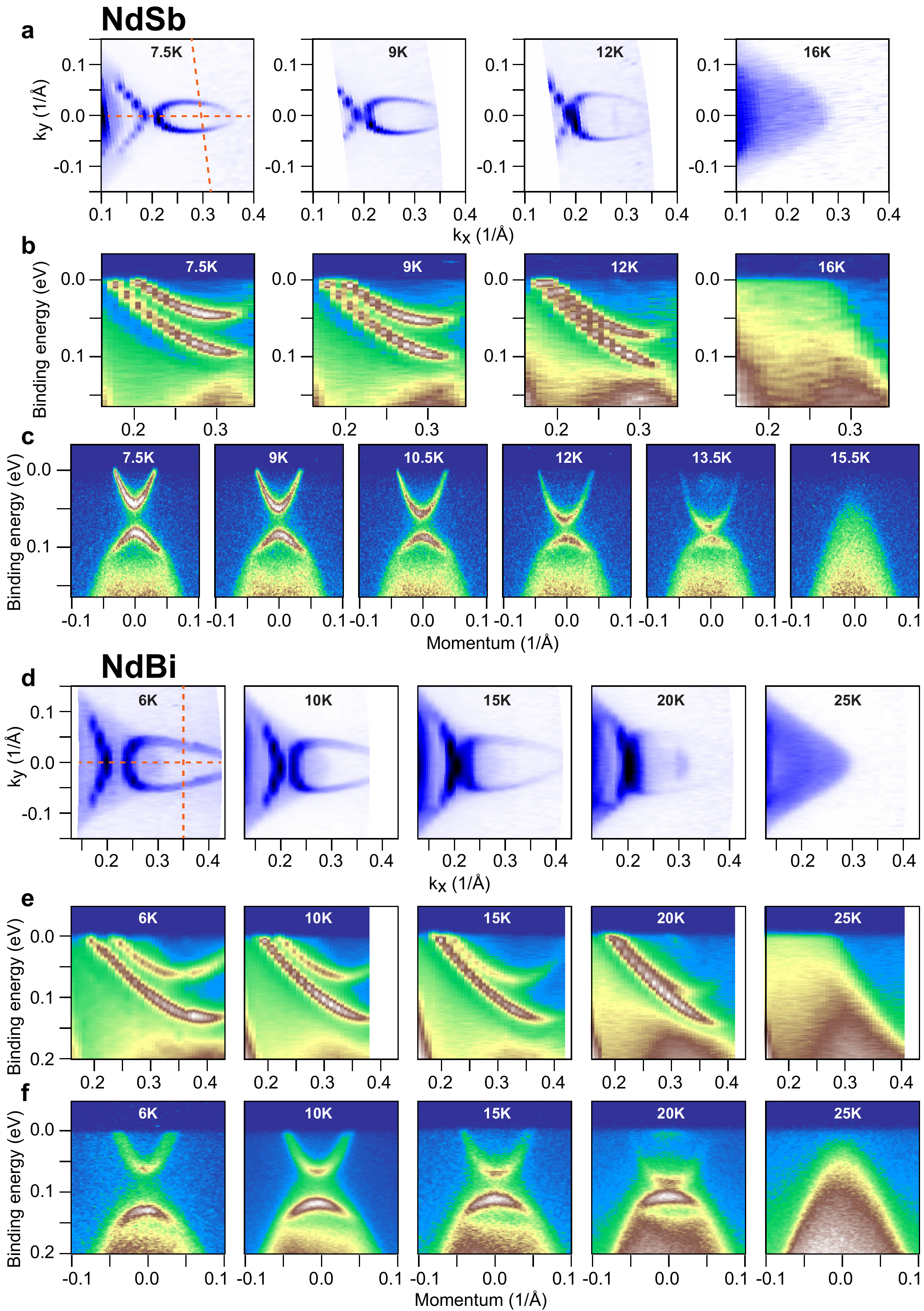}%
	\caption{{\bf $|$ Temperature evolution of the surface state dispersion for NdBi and NdSb.}
    {\bf a}, Fermi surface map of NdSb at at four temperatures below and one above $T_N$.
    {\bf b} and {\bf c}, Band dispersion along directions marked with the horizontal and vertical dashed line in {\bf a} respectively.
    {\bf d}, Fermi surface map of NdBi at at five temperatures below and one above $T_N$.
    {\bf e} and {\bf f}, Band dispersion along directions marked with the horizontal and vertical dashed line in {\bf d} respectively.
	\label{fig:fig4}}
\end{figure*}

\begin{figure*}[bt]
	\includegraphics[width=6 in]{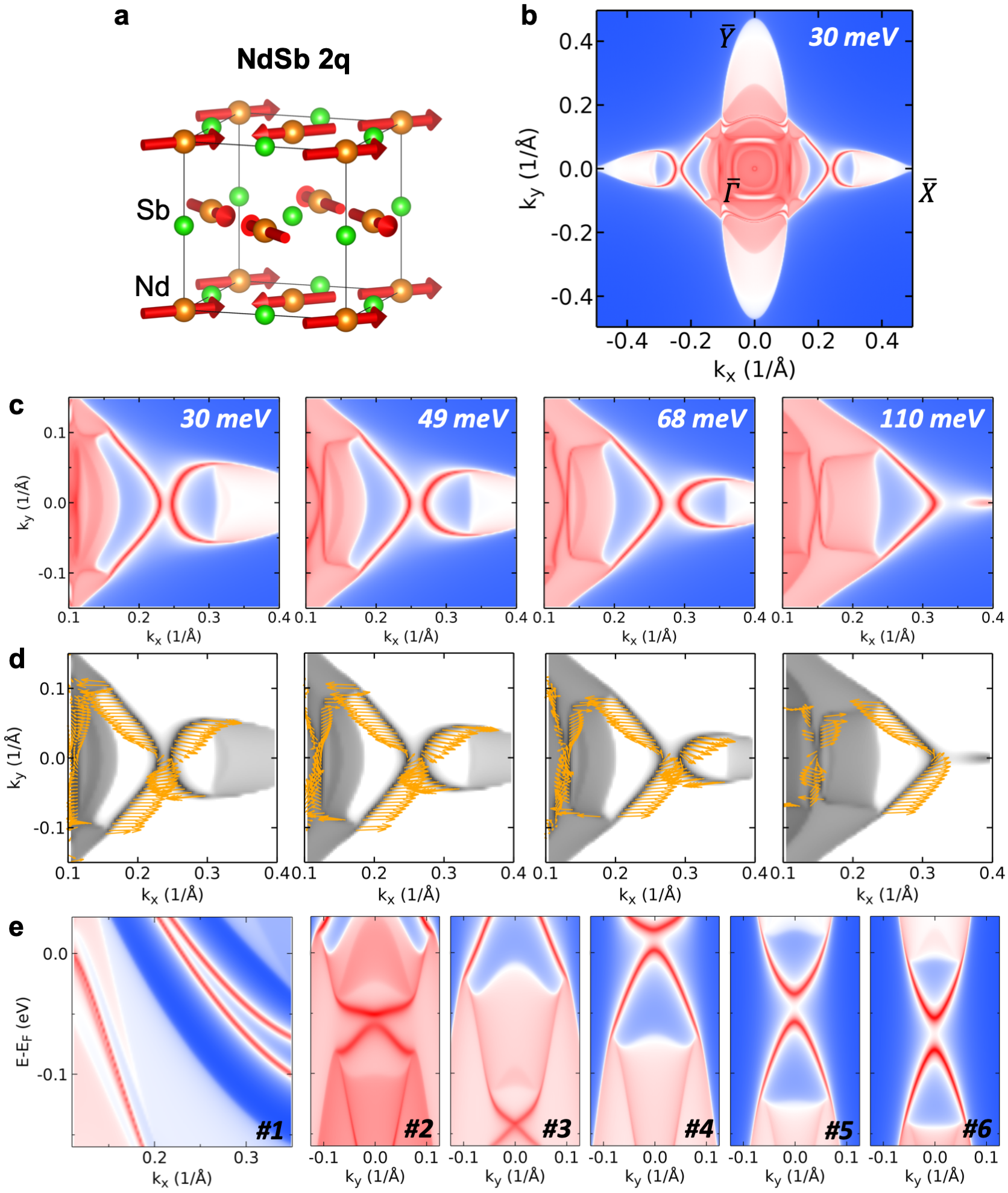}%
	\caption*{Fig. 5. Results of DFT calculations.
	{\bf a} Magnetic structure in the case of 2$q$ NdSb. The arrows show the directions of Nd magnetic moments. The (001) direction corresponds to top surface while (100) and (010) to side surfaces.
	{\bf b} Calculated 2D surface Fermi surface of NdSb in AFM2q ordered state at (010) surface.
	{\bf c} Constant energy plots of small region from panel (b) where Fermi arcs exist.
	{\bf d} the spin textures for constant energy contours shown in (c) indicated by the orange arrows.
	{\bf e} Band dispersion with a shift of 30 meV along the $\Gamma-X$ direction (cut \#1) and cuts \#2-6 at $k_y$=0.14. 0.18, 0.23, 0.29 and 0.32 corresponding to ARPES data in Fig.2b.
	\label{fig:xrd}}
\end{figure*}

\newpage

\clearpage

{\bf Supplementary Information for ``Rare-earth monopnoctides - family of antiferromagnets hosting magnetic Fermi arcs."}

\bigskip
\bigskip

{\bf Additional data: CeBi}

In the left column of Fig.~S1a, we show cuts from Fig.~3c in a wider momentum range and several more cuts obtained from the same data set. For comparison in the right column, we plot corresponding cuts obtained from the data-set measured in the PM state. The presence of two SS dispersions is seen on both sides of the low-temperature plots. The marginal asymmetry of these plots is caused by a slight misalignment of the sample. Different shapes of the lower and upper SS dispersion indicate that the observed splitting is real but not a result of detecting signal from two sample crystallites. This conclusion is confirmed by observing two SS dispersions in the data obtained from two other samples: Fig.~S1b and Fig.~S2a.

Besides SS dispersions (marked with white arrows), low-temperature plots display another sharp feature (marked with red arrows). However, in contrast to the topological SS, this dispersion is also seen in the data measured in the PM phase (second column of Fig.~S1b. It is formed by the bulk states and form apart of the outer pocket around $\Gamma$-point (particularly a part of it marked with a red arrow in Fig.~1c and f).

In Fig.~S2a and b, we show the spectrum from Fig.~3e and a corresponding spectrum measured in the PM state. In the PM state, we observe only one wide blob of intensity associated with bulk states. In the AFM state, two additional features appear. This is even better seen from the EDCs taken throw the band center (see Fig.~S2c). Here, the 26.5K curve shows a smooth background of the bulk states, and the 13.5K curve shows two pronouns peaks on this background. This result, together with the similarity of Fig.~S2a and Fig.~4c (20~K), confirms that these features are associated with two topological SS.

\begin{figure*}[tb]
	\includegraphics[width=6 in]{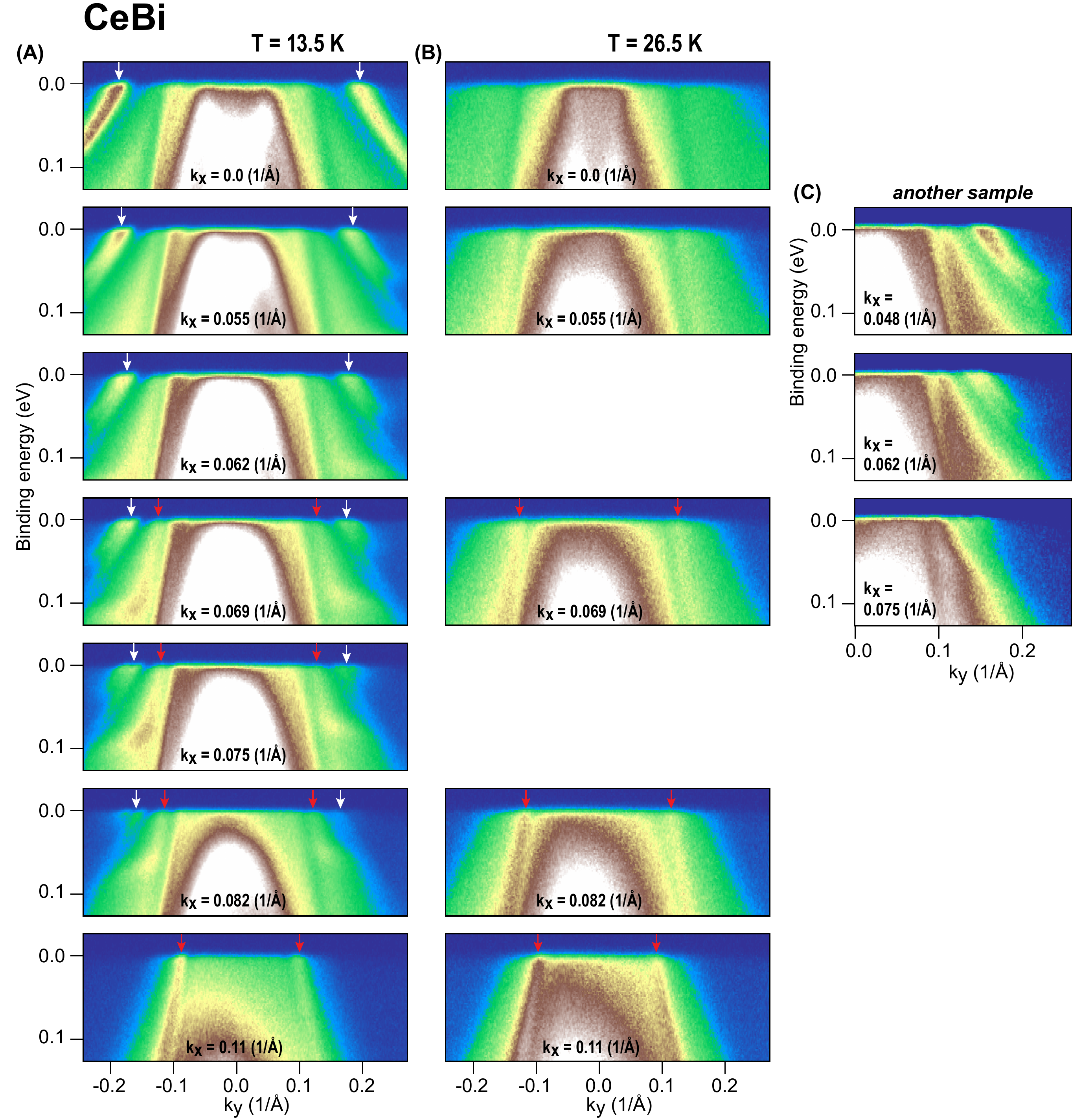}%
	\caption*{Supplementary Fig. S1. Band dispersion in the AFM and PM states of CeBi. 
	{\bf a} Band dispersions along several cuts parallel $\Gamma-X$ direction measured in the AFM state. Here, white and red arrows indicate the location of the surface state and bulk dispersions, respectively.
	{\bf b} The same band dispersions as in (a) but measured in the PM state.
	{\bf c} Band dispersions measured in the AFM state from another sample.
	\label{fig:xrd}}
\end{figure*}

\begin{figure*}[tb]
	\includegraphics[width=6.5 in]{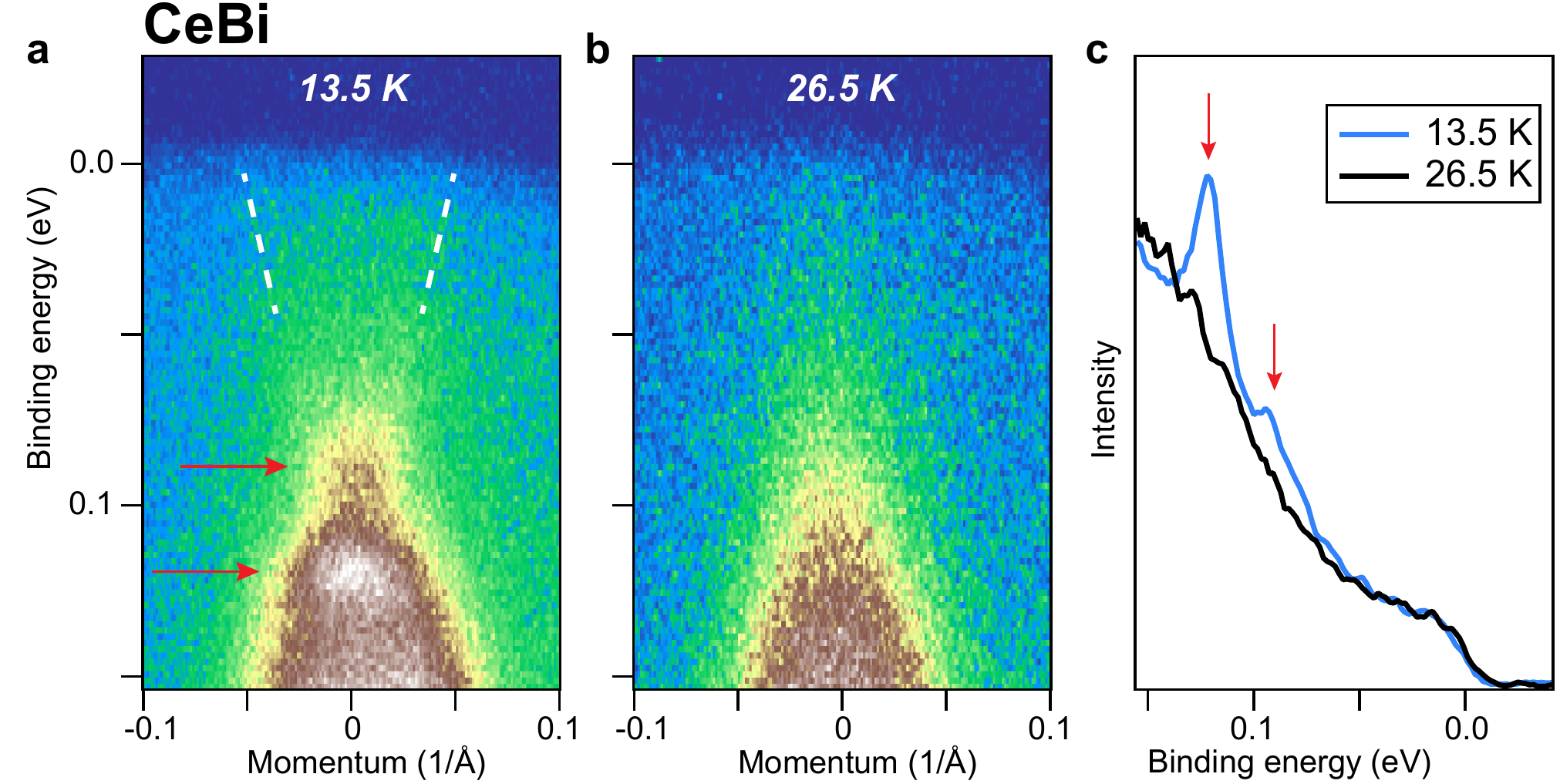}%
	\caption*{Supplementary Fig. S2. Band dispersion below and above T$_N$ in CeBi. 
	{\bf a} Spectrum measured in the AFM state (the same as Fig.3e).
	{\bf b} The same spectrum as in (a) but measured in the PM state.
	{\bf c} EDCs taken throw the band center from (a) and (b)
	\label{fig:xrd}}
\end{figure*}

\clearpage

{\bf Domains in NdSb and NdBi}

In Fig.S4, we show one more Fermi Surface map of NdSb measured in the AFM state. In this map, one can see the same arc features and ellipse pockets as in Fig.1d. However, in contrast to Fig.1d, here, they are present in both directions. Since both maps were measured with the same (linearly vertical polarised) light, almost complete absents of the SS features in one direction in Fig.1d can not be a result of the matrix elements, but it is an accurate representation of the Fermi Surface. This is a result of measuring photoemission from a single domain, when Fig.S4 is a supposition of signal from domains of two orientations. Also from this result we see that one domain orientation is not dominant over the whole sample surface of NdSb.

The presence of the SS features in both directions makes Fig.S4 similar to Fig.1e and other available Fermi Surface maps of NdBi$^{24}$. This similarity indicates that the Fermi Surface maps of NdBi are also a supposition of signal from domains of two orientations. The reason why a single domain of NdBi has not been seen in ARPES can be a small domain size in this material (smaller than the beam size), and as a result, one always gets a superposition of signal from multiple domains.

\begin{figure*}[bt]
	\includegraphics[width=3 in]{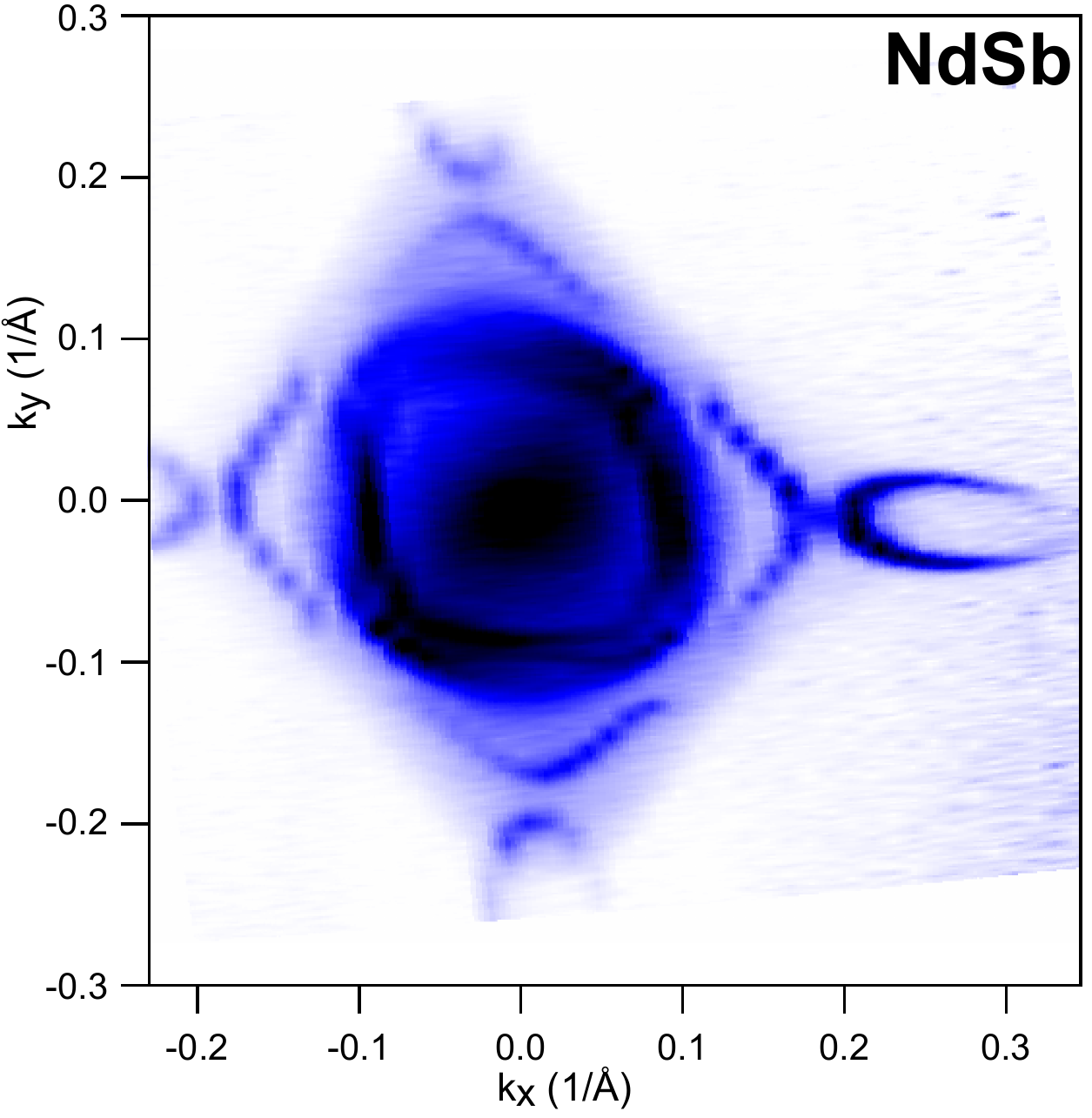}%
	\caption*{Supplementary Fig. S3. Fermi Surface map of NdSb that is a superposition of signals from domains of different orientations.
	\label{fig:xrd}}
\end{figure*}
\clearpage





\clearpage

{\bf Additional data: SmBi}

In Fig.S4b, we show two orthogonal cuts that represent $\Gamma-X$ direction obtained from the same data set as the Fermi surface map in Fig.1f. In contrast to similar cuts for three other materials (see Fig.2 b\#1 and d\#1 and Fig. 3d), these cuts do not show the presence of any signs of sharp surface states. A spectrum measured along $\Gamma-X$ of another sample down to 6.2K (Fig.S4c) also does not indicate the presence of these states and does not qualitative change with increasing temperature above $T_N$.

\begin{figure*}[bt]
	\includegraphics[width=6 in]{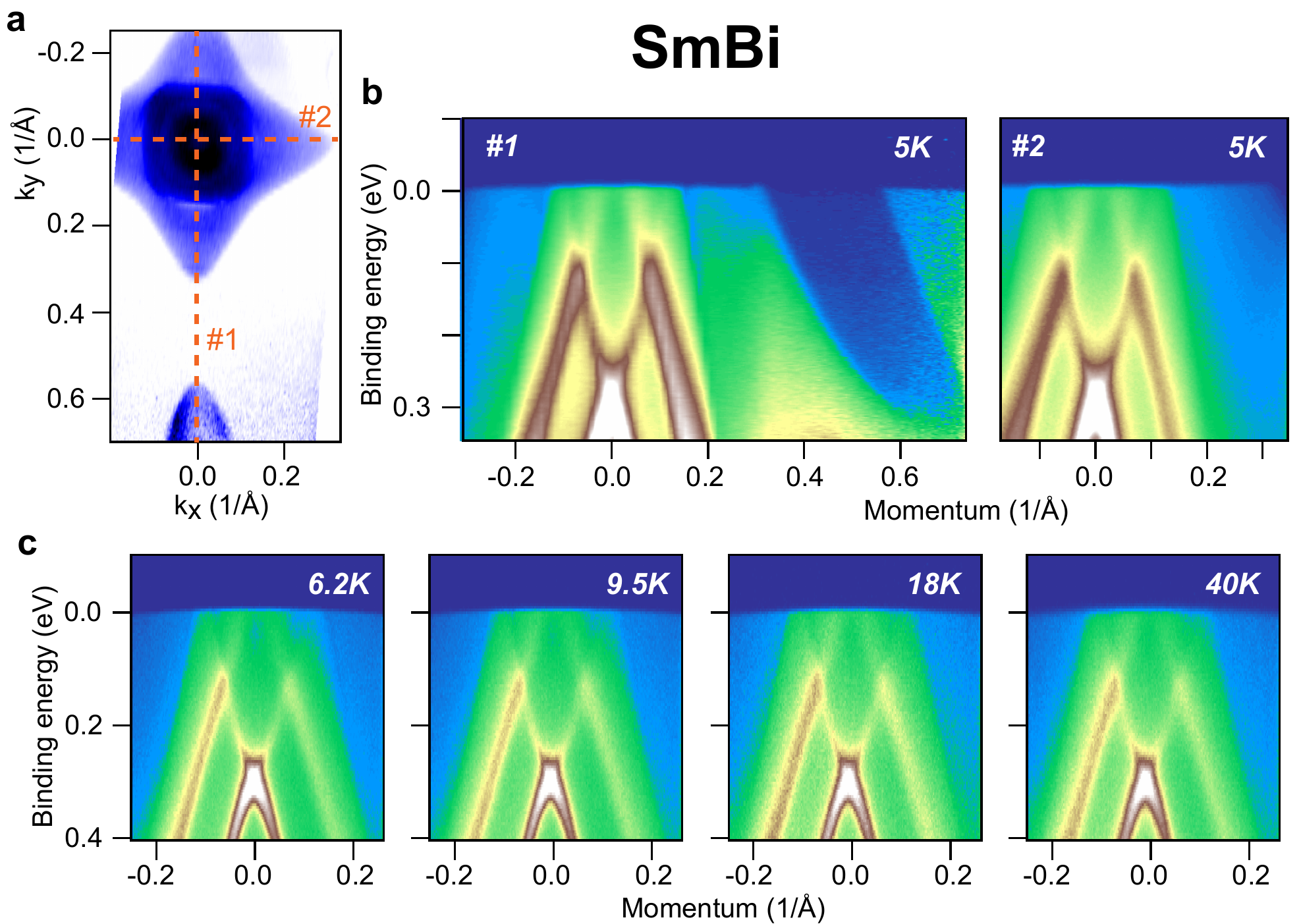}%
	\caption*{Supplementary Fig. S4. Additional data from SmBi.
	{\bf a} Fermi surface of SmBi in the AFM state (the same as Fig.1f).
	{\bf b} Band dispersions along the directions marked with the dashed line in (a).
	{\bf c} Temperature dependence of the band dispersions along $\Gamma-X$ direction.
	\label{fig:xrd}}
\end{figure*}

\end{document}